\documentclass[aps,prd,onecolumn,superscriptaddress,tighten,nofootinbib]{revtex4}
\usepackage{amsfonts,amssymb,latexsym,amsmath} 
\usepackage[colorlinks,citecolor=blue,urlcolor=blue,linkcolor=blue]{hyperref}
\usepackage{bm}
\usepackage{hhline}
\usepackage{longtable}
\usepackage{array} 
\usepackage{color}
\usepackage{ifthen}
\usepackage{epsfig}
 \usepackage{multirow}
\usepackage{enumerate}
\usepackage{dsfont}
\usepackage{psfrag}
\usepackage{float}
\usepackage{epstopdf}
\usepackage{mathtools,slashed}
\usepackage{xcolor,booktabs}
\usepackage{comment}
\usepackage{lipsum}

\newcommand{\RNum}[1]{\uppercase\expandafter{\romannumeral #1\relax}}

\newcommand{\be}{\begin{equation}}
 \newcommand{\ee}{\end{equation}}
\newcommand{\ba}{\begin{array}{c}}
 \newcommand{\ea}{\end{array}}
\newcommand{\bea}{\begin{eqnarray}}
 \newcommand{\eea}{\end{eqnarray}}

\allowdisplaybreaks

\begin{document}
\title{The radiative decays of the singly heavy baryons in chiral perturbation theory}

\author{Guang-Juan Wang}\email{wgj@pku.edu.cn}
\affiliation{School of Physics and State Key Laboratory of Nuclear
Physics and Technology, Peking University, Beijing 100871, China}

\author{Lu Meng}\email{lmeng@pku.edu.cn}
\affiliation{School of Physics and State Key Laboratory of Nuclear
Physics and Technology, Peking University, Beijing 100871, China}

\author{Shi-Lin Zhu}\email{zhusl@pku.edu.cn}\affiliation{School of Physics and State Key Laboratory of Nuclear Physics and Technology, Peking University, Beijing 100871, China}\affiliation{Collaborative Innovation Center of Quantum Matter, Beijing 100871, China}

\begin{abstract}
In the framework of the heavy baryon chiral perturbation theory (HBChPT), we calculate the radiative decay amplitudes of the singly heavy baryons up to the next-to-next-to-leading order (NNLO). In the numerical analysis, we adopt the heavy quark symmetry to relate some  low energy constants (LECs) with those LECs in the calculation of the magnetic moments. We use the results from the lattice QCD simulation as input. With a set of unified LECs, we obtain the numerical (transition) magnetic moments and radiative decay widths. We give the numerical results  for the spin-$1\over 2$ sextet to the spin-$1\over 2$ antitriplet up to the next-to-leading order (NLO). The nonvanishing $\Gamma(\Xi_c^{'0} \rightarrow \Xi^0_c\gamma)$ and $\Gamma(\Xi_c^{*0} \rightarrow \Xi^0_c \gamma)$ solely arise from the U-spin symmetry breaking, and do not depend on the lattice QCD inputs up to NLO. We also systematically give the numerical analysis of the magnetic moments of the spin-$1\over 2$, spin-$3\over 2$ sextet and their radiative decay widths up to NNLO. In the heavy quark limit, the radiative decays between the sextet states happen through the magnetic dipole (M1) transitions, while the electric quadrupole (E2) transition does not contribute. We also extend the same analysis to the single bottom baryons.

 \end{abstract}


\maketitle

\section{Introduction}
A heavy baryon contains two light quarks and a heavy quark. In the SU(3) flavor symmetry, the two light quarks form a diquark in the antisymmetric $\bar 3_f$ or the symmetric $6_f$ representation. Constrained by the Fermi-Dirac statistics, the $J^P$ of the diquark is $0^+$ or $1^+$, respectively. Then the diquark and the heavy quark are combined to form the heavy baryon. 
For the ground antitriplet, the $J^P$ is $\frac{1}{2}^+$. For the ground sextet, the $J^P$ is $\frac{1}{2}^+$ or $\frac{3}{2}^+$. In the following, we use $\psi_{\bar 3}$, $\psi_6$, and $\psi^{\mu}_{6^*}$ to denote the spin-$1\over 2$ antitriplet, spin-$1\over 2$, and spin-$3\over 2$ sextet, respectively.

For the transitions $\psi_6\rightarrow \psi_{\bar 3}\text{ and }\psi^{\mu}_{6^*}\rightarrow\psi_{6}$,
 the radiative decays are quite important, since some strong decay channels are forbidden by the phase space. So far, the BaBar and Belle Collaborations have observed three radiative decay processes: $\Omega_c^*\rightarrow \Omega_c \gamma$ \cite{Aubert:2006je,Solovieva:2008fw}, $\Xi_c^{'+} \rightarrow \Xi_c^{+} \gamma$ and $\Xi_c^{'0} \rightarrow \Xi_c^0 \gamma$~\cite{Jessop:1998wt,Aubert:2006rv,Yelton:2016fqw}. More observations are expected at the BESIII, Belle II, LHCb and other collaborations in the future. 

 The radiative decay processes are good platforms for studying the electromagnetic properties, which are important to reveal the inner structures of the heavy baryons.
 In literature, theorists used many different models to study the radiative decays. In Refs.~\cite{Bahtiyar:2015sga,Bahtiyar:2016dom}, the authors studied the decay widths and electromagnetic form factors of the processes $\Omega_c^*\rightarrow \Omega_c \gamma $ and $\Xi_c' \rightarrow \Xi_c \gamma$ using the lattice QCD simulation. In Ref.~\cite{Cheng:1992xi}, the authors constructed the chiral Lagrangains for the heavy baryons incorporating the heavy quark symmetry and studied the radiative decays of the heavy baryons and mesons. Later, the authors in Refs.~\cite{Cho:1994vg,Savage:1994wa,Banuls:1999br,Tiburzi:2004mv} investigated the electromagnetic properties of the heavy baryons in the heavy hadron chiral perturbation theory. In Ref.~\cite{Jiang:2015xqa}, Jiang {\it et al.} calculated the electromagnetic decay widths of the heavy baryons up to the next-to-leading order (NLO) in the heavy baryon chiral perturbation theory (HBChPT). They found that the neutral radiative decay channels, e.g. $\Xi_c^{'0} \rightarrow \Xi_c^0 \gamma$ and $\Xi_c^{*0} \rightarrow \Xi_c^{'0} \gamma$, are suppressed due to the U-spin symmetry. Besides the lattice QCD and the effective field theory, theorists also studied the radiative decays with other phenomenological models: the heavy quark symmetry~\cite{Tawfiq:1999cf}, the light cone QCD sum rule formalism~\cite{Wang:2009ic, Agamaliev:2016fou,Aliev:2011bm,Zhu:1997as,Zhu:1998ih}, the bag model~\cite{Simonis:2018rld,Bernotas:2013eia}, the nonrelativistic quark model~\cite{Majethiya:2009vx}, the relativistic three quark model~\cite{Ivanov:1999bk} and other various quark models~\cite{JuliaDiaz:2004vh, Faessler:2006ft, Albertus:2006ya,Dey:1994qi, Sharma:2010vv, Barik:1984tq,Wang:2017kfr}.

The chiral perturbation theory is firstly used to study the properties of the pseudoscalar mesons~\cite{Gasser:1983yg,Gasser:1984gg,Weinberg:1978kz, Bijnens:1995yn}. It has a self-consistent power counting law which is in terms of the small momentum (mass) of the pseudoscalar mesons. When it is extended to the baryons, the mass of a baryon, which is at the same order as the chiral symmetry breaking scale in the chiral limit, breaks the consistent power counting~\cite{Gasser:1987rb}. To solve this problem, the heavy baryon chiral perturbation theory (HBChPT) is developed~\cite{Jenkins:1990jv,Bernard:1992qa,Ecker:1995rk,Muller:1996vy}. In this scheme, the baryon field is decomposed into the light and heavy components. The heavy component can be integrated out in the low energy region and the large mass of the baryon is eliminated. Now, the power counting law recovers and the expansion is in terms of the residue momentum of the baryons and the momentum (mass) of the pseudoscalar mesons.

So far, the amplitudes of the radiative decays are calculated up to NLO using the effective theory \cite{Cheng:1992xi,Cho:1994vg,Savage:1994wa,Banuls:1999br,Tiburzi:2004mv,Jiang:2015xqa}. In this work, we systematically derive the radiative decay amplitudes up to the next-to-next-to-leading order (NNLO) in HBChPT. Many low energy coefficients (LECs) are involved in the analytical expressions. Some of them also appear in the magnetic moments up to NNLO. In this work, we will use the data of the magnetic moments and the radiative decay widths from the lattice QCD simulations as input to obtain the numerical results. In the numerical analysis, we adopt the heavy quark symmetry to reduce the number of the LECs~\cite{Wang:2018gpl,Meng:2018gan}. We give the final results of (transition) magnetic moments, radiative decay widths in a group of unified LECs.
 
The paper is arranged as follows. In Section~\ref{sec1}, we derive the expressions of the decay widths using the form factors from the electromagnetic multipole expansion. In Section~\ref{sec2}, we present the effective Lagrangians that contribute to the radiative decays up to NNLO. In Section~\ref{sec3}, we derive the analytical expressions of the decay amplitudes up to NNLO. In Section~\ref{sec4}, we construct the Lagrangians in the heavy quark limit and reduce the number of the LECs using the heavy quark symmetry. 
In Section~\ref{sec5}, we use the data from the lattice QCD simulation as input to calculate the LECs. Then, we obtain the numerical results of the (transition) magnetic moments, the M1 transition form factors and the decay widths of the charmed baryons up to NNLO. In Section~\ref{sec6}, we extend the calculations to the bottom baryons. Finally, we compare our results with those from other models and give a brief summary in~\ref{sec7}. In Appendix~\ref{app:mm}, we give the magnetic moments of the spin-$1\over 2$ and spin-$3\over 2$ sextet as by-product. In Appendix~\ref{QM}, we give some quark model results. In Appendix~\ref{loop}, we list the details of the loop integrals.

\section{The radiative decay width}\label{sec1}
In the SU(3) flavor symmetry, the explicit matrix forms of the spin-$\frac{1}{2}$
antitriplet, spin-$\frac{1}{2}$, and spin-$\frac{3}{2}$ sextet fields are
\begin{eqnarray}
\quad\psi_{{\bar{3}}}=\left(\begin{array}{ccc}
0 & \Lambda_{c}^{+} & \Xi_{c}^{+}\\
-\Lambda_{c}^{+} & 0 & \Xi_{c}^{0}\\
-\Xi_{c}^{+} & -\Xi_{c}^{0} & 0
\end{array}\right),\quad\psi_{{6}}=\left(\begin{array}{ccc}
\Sigma_{c}^{++} & \frac{\Sigma_{c}^{+}}{\sqrt{2}} & \frac{\Xi_{c}^{\prime+}}{\sqrt{2}}\\
\frac{\Sigma_{c}^{+}}{\sqrt{2}} & \Sigma_{c}^{0} & \frac{\Xi_{c}^{\prime0}}{\sqrt{2}}\\
\frac{\Xi_{c}^{\prime+}}{\sqrt{2}} & \frac{\Xi_{c}^{\prime0}}{\sqrt{2}} & \Omega_{c}^{0}
\end{array}\right),\quad\psi_{6^{*}}^{\mu}=\left(\begin{array}{ccc}
\Sigma_{c}^{*++} & \frac{\Sigma_{c}^{*+}}{\sqrt{2}} & \frac{\Xi_{c}^{*+}}{\sqrt{2}}\\
\frac{\Sigma_{c}^{*+}}{\sqrt{2}} & \Sigma_{c}^{*0} & \frac{\Xi_{c}^{*0}}{\sqrt{2}}\\
\frac{\Xi_{c}^{*+}}{\sqrt{2}} & \frac{\Xi_{c}^{*0}}{\sqrt{2}} & \Omega_{c}^{*0}
\end{array}\right)^{\mu}.
 \end{eqnarray} 
In the following, we calculate the decay widths of the transitions: $\psi_6\rightarrow \psi_{\bar 3} \gamma $, $\psi^{\mu}_{6^*}\rightarrow \psi_{\bar 3} \gamma $, and $\psi^{\mu}_{6^*}\rightarrow \psi_{6} \gamma $ in the HBChPT scheme, respectively. 
\subsection{The radiative transition: spin-$\frac{1}{2}+\gamma\rightarrow$ spin-$\frac{1}{2}$}
For the radiative decay from the spin-$1\over 2$ sextet to the spin-$1\over 2$ antitriplet, the decay amplitude reads~\cite{Leinweber:1990dv,Kubis:2000aa},
\begin{eqnarray}\label{r1}
i\mathcal{M}&=&-ie\epsilon^{\mu}\langle {\psi}(p')|j_{\mu}|\psi(p)\rangle \nonumber \\
 &=&-ie\epsilon^{\mu}\bar{u}(p')\left[\left(\gamma_{\mu}-\frac{\delta}{q^2}q_\mu\right)F_{1}(q^2)+\frac{i\sigma_{\mu\nu}q^{\nu}}{M+M'}F_{2}(q^2)\right]u(p),
 \end{eqnarray} 
 where $\epsilon^{\mu}$ is the polarization vector of the photon. $j_{\mu}$ is the electromagnetic current. $u(p)$ is the heavy baryon field with momentum $p$. The momentum transformed is $q=p'-p$. $F_{1,2}$ are the form factors with $q^2$ as the variable. $M$ and $M'$ are the initial and final heavy baryon masses, respectively. $\delta=M'-M$ is the mass difference.

In HBChPT, one decomposes the momentum of a heavy baryon as 
\begin{eqnarray}
p^\mu=Mv^\mu+k^\mu,
\end{eqnarray}
where $k^\mu$ is a small residue momentum. $v^\mu$ is the velocity of the heavy baryon and satisfies $v^2=1$. The field $\psi$ is then decomposed into the ``light'' component $B(p)$
and ``heavy'' component $H(p)$ as follows,

\begin{eqnarray}
B(p)=e^{iMv\cdot x}\frac{1+\slashed v}{2}\psi,~~~H(p)=e^{iMv\cdot x}\frac{1-\slashed v}{2}\psi.
 \end{eqnarray} 
In the low energy region, one can integrate out the heavy component $H(p)$ and obtain Lagrangians in the nonrelativistic limit. At this time, the electromagnetic matrix element in Eq.~(\ref{r1}) is written as, 

\begin{eqnarray}
 \langle\psi(p')|j_{\mu}|\psi(p)\rangle &=&e\bar{B}(p')\left[\left(v_{\mu}-\frac{\delta }{q^2}q_\mu \right)G_{E}(q^{2})+\frac{2[S^{\mu},S^{\nu}]q_{\nu}}{M+M'}G_{M}(q^{2})\right]B(p)\text{,}\\
 G_{E}(q^{2})&=&F_{1}(q^{2})+\frac{q^{2}}{2(M+M')^{2}}F_{2}(q^{2}),\\
 G_{M}(q^{2})&=&F_{2}(q^{2})+F_{1}(q^{2}),
 \end{eqnarray}
where $S^{\mu}$ is the Pauli-Lubanski operator $\frac{i}{2}\gamma^{5}\sigma^{\mu\nu}v_{\nu}$. $G_E$ and $G_M$ are the charge (E0) and the magnetic dipole (M1) form factors, respectively. When $q^2=0$, $G_E(0) = F_1(0)\approx 0$ because of the orthogonality of the initial and final states. Then, the $G_M(0)\approx F_2(0)$ and the decay width is expressed by the magnetic form factor $G_M(0)$ as~\cite{Bahtiyar:2016dom,Cheng:1992xi}
\begin{eqnarray}
 \Gamma=\frac{4\alpha |\mathbf{p}_{\gamma}|^{3}}{(M+M')^{2}}|G_{M}(0)|^{2},
 \end{eqnarray}
 where $\alpha=\frac{e^2}{4\pi}\approx{1\over 137}$ is the fine-structure constant, $\mathbf{p}_{\gamma}$ is the momentum of the photon in the central mass system of the initial state,
 \begin{eqnarray}
 |\mathbf{p}_{\gamma}|=\frac{M'^2-M^2}{2M'}.
 \end{eqnarray}

\subsection{The radiative transition:  spin-$\frac{3}{2}\rightarrow$ spin-$ \frac{1}{2}+\gamma$}
To calculate the decay amplitude of the radiative transition $  \psi^{\mu}_{6^*}\rightarrow \psi_{\bar 3, 6}+\gamma$, we introduce the multipole expansion of the electromagnetic current matrix element~\cite{Faessler:2006ky,Jones:1972ky}
\begin{eqnarray}
&\langle\psi_{6^{*}}|j_{\mu}|\psi\rangle=e\bar{u}^{\rho}(p')\Gamma_{\text{\ensuremath{\rho}}\mu}u(p),\nonumber 
\end{eqnarray}
with
\begin{eqnarray}
\Gamma_{\rho\mu} =G_{1}(q^{2})(q_{\rho}\gamma_{\mu}-\slashed{q} g_{\rho\mu})\gamma_{5}+G_{2}(q^{2})(q_{\rho}p'_{\mu}-q\cdot p'g_{\rho\mu})\gamma_{5}+G_{3}(q^{2})(q_{\rho}q_{\mu}-q^{2}g_{\rho\mu})\gamma_{5}.
 \end{eqnarray}
where the factors $G_{1,2,3}$ are functions of $q^{2}$. In the HBChPT scheme, the nonrelativistic form of the $\Gamma_{\rho\mu} $ is
\begin{eqnarray}
\Gamma_{\rho\mu} =2G_{1}(q^{2})(q_{\rho}S_{\mu}-q\cdot S g_{\rho\mu})+G_{2}(q^{2}){2M'\over{M+M'}}(q_{\rho}v_{\mu}-q\cdot v g_{\rho\mu})q\cdot S,
 \end{eqnarray}
where we have omitted the $G_3$ term since it does not contribute when the photon is on-shell. The spin-$3\over2$ state decays into the spin-$1\over 2$ state through the M1 and E2 transitions. The corresponding magnetic dipole form factor $G_{M1}$ and
electric quadrupole form factor $G_{E2}$ can be constructed using $G_{1,2,3}$ as follows, 

\begin{eqnarray}\label{tr13}
&&G_{M1}(q^2) =\frac{1}{4}\left[G_{1}\frac{M_{+}(3M'+M)-q^{2}}{M'}+G_{2}(M_{+}M_{-}-q^{2})+2(G_{3}+G_{2})q^{2}\right],\nonumber \\
&&G_{E2} (q^2)=\frac{1}{4}\left[G_{1}\frac{M_{+}M_{-}+q^{2}}{M'}+G_{2}(M_{+}M_{-}-q^{2})+2(G_{2}+G_{3})q^{2}\right],
 \end{eqnarray}
where the $M_{\pm}=M'\pm M$. Since $M$ and $M'$ are roughly the same, $M_{-}$ is nearly $0$. When $q^2=0$, we obtain $ \frac{G_{E2}(0)}{G_{M1}(0)}=\frac{M_-}{M} (\frac{1}{4}+\frac{G_2}{ 4G_1})$, which indicates the $G_{E2}(0)$ is much smaller than $G_{M1}(0)$.
 
The transition magnetic moment is defined as 
\begin{eqnarray}\label{tr14}
\mu=\sqrt{\frac{2}{3}}\frac{G_{M1}(0)}{2M}.
\end{eqnarray}
With the $G_{M1} $ and $G_{E2} $, the helicity amplitudes are defined as,
\begin{eqnarray}\label{tr15}
A_{3/2}(q^2) =-\sqrt{\frac{\pi\alpha\omega}{2M^{2}}}[G_{M1}(q^2) +G_{E2}(q^2) ],\\
A_{1/2} (q^2)=-\sqrt{\frac{\pi\alpha\omega}{6M^{2}}}[G_{M1}(q^2) -3G_{E2}(q^2) ],
 \end{eqnarray}
with 
\begin{eqnarray}
\omega=\frac{M^{'2}-M^{2}+q^{2}}{2M'}.
 \end{eqnarray}
The radiative decay width reads,
\begin{eqnarray}\label{tr16}
\Gamma=\frac{M M'}{8\pi}\left(1-\frac{M^{2}}{M'^{2}}\right)^{2}\left(A_{3/2}^{2}(0)+A_{1/2}^{2}(0)\right).
\end{eqnarray}

\section{The Lagrangian}\label{sec2}
\begin{figure}[tb]
 \centering
 \includegraphics[scale=0.6]{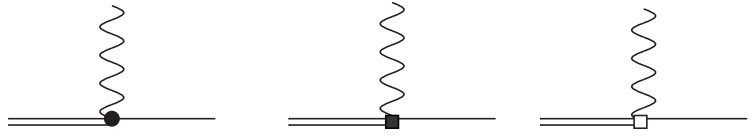}
 \caption{The tree level diagrams that contribute to the radiative decays. The solid circle, solid square and the square denote the vertices at $\mathcal O(p^2)$, $\mathcal O(p^3)$, and $\mathcal O(p^4)$, respectively. The single and double lines denote the spin-$1\over 2$ and spin-$3\over 2$ heavy baryons, respectively.}\label{tree}
 \end{figure}

\begin{figure}[htbp] 
\centering\includegraphics[width=6in]{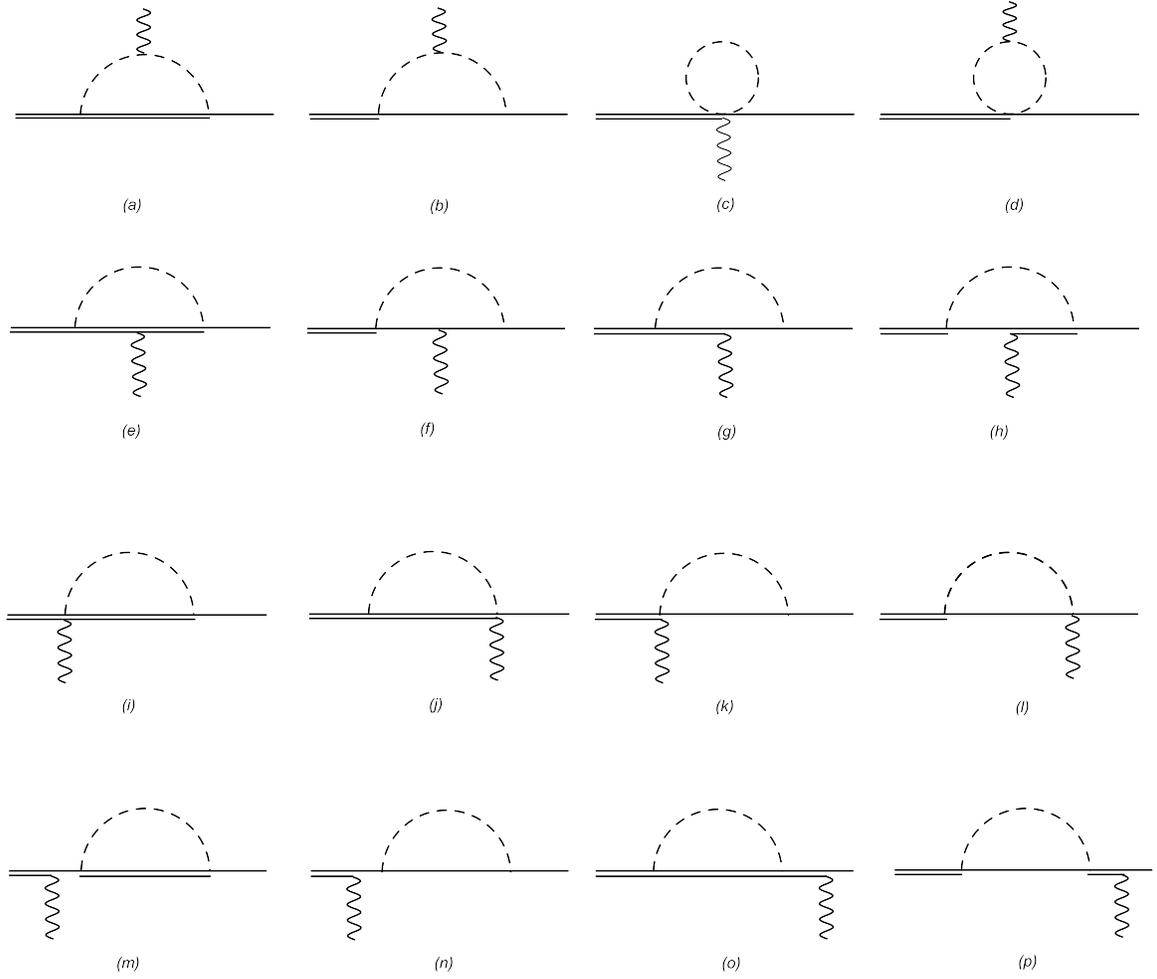} 
\caption{The one loop diagrams which contribute to the radiative transitions up to $\mathcal O(p^4)$. The single and double lines denote the spin-$1\over 2$ and spin-$3\over 2$ heavy baryons, respectively.}\label{allloop} 
\end{figure}

We list the tree and loop diagrams that contribute to the radiative decay amplitudes up to $\mathcal O (p^4)$ in Fig.~\ref{tree} and Fig.~\ref{allloop}, respectively. The diagram with chiral dimension $D_{\chi}$ contributes to the $\mathcal {O}(p^{D_{\chi}})$ radiative decay amplitude and $\mathcal {O}(p^{D_{\chi}-1})$ transition magnetic moment. In the following, we list the Lagrangians involved in this work. 

 The leading-order Lagrangian for the pseudoscalar meson interaction reads
 \begin{eqnarray}
\mathcal{L}_{\phi}^{(2)}=\frac{F_{\phi}^{2}}{4}\text{Tr}[\triangledown_{\mu}U\triangledown^{\mu}U^{\dagger}],
 \end{eqnarray}
with 
\begin{eqnarray}
&U=\xi^{2}=e^{\frac{i\phi}{F_{\phi}}},\quad\triangledown_{\mu}U=\partial_{\mu}U+ieA_{\mu}[Q,U],\nonumber\\
&\phi=\sqrt{2}\left(\begin{array}{ccc}
\frac{\pi^{0}}{\sqrt{2}}+\frac{\eta}{\sqrt{6}} & \pi^{+} & K^{+}\\
\pi^{-} & -\frac{\pi^{0}}{\sqrt{2}}+\frac{\eta}{\sqrt{6}} & K^{0}\\
K^{-} & \overline{K}^{0} & -\frac{2}{\sqrt{6}}\eta
\end{array}\right)~\text{and}~Q=\left(\begin{array}{ccc}
\frac{2}{3} &0 &0 \\
0 &-\frac{1}{3} & 0\\
0 &0 &-\frac{1}{3}
\end{array}\right),
 \end{eqnarray}
where $A_{\mu}$ is the photon field, $Q$ is the charge matrix of the light quarks. We use the $m^\phi$ and $F^\phi$ to denote the masses and the decay constants of the mesons, respectively. Their values are~\cite{Gasser:1984gg}
\begin{eqnarray}
&m_\pi=137~ \text{MeV},~~m_K= 496~\text{MeV},~~m_\eta= 548~\text{MeV},\nonumber \\
&F_\pi=92.4~\text{MeV},~~F_K=113~\text{MeV},~~F_\eta=116~\text{MeV}.
 \end{eqnarray}

The leading-order meson-baryon Lagrangian $\mathcal L^{(1)}_{B\phi}$ reads
\begin{eqnarray}
\mathcal{L}_{B\phi}^{(1)}&=&\frac{1}{2}\text{Tr}[\bar{\psi}_{\bar{3}}(i\slashed D-M_{\bar{3}})\psi_{\bar{3}}]+\text{Tr}[\bar{\psi}_{6}(i\slashed D-M_{6})\psi_{6}]\nonumber\\
&&+\text{Tr}[\bar{\psi}_{6^{*}}^{\mu}(-g_{\mu\nu}(i\slashed D-M_{6^{*}})+i(\gamma_{\mu}D_{\nu}+\gamma_{\nu}D_{\mu})-\gamma_{\mu}(i\slashed D+M_{6^{*}})\gamma_{\nu})\psi_{6^{*}}^{\nu}]\nonumber\\
 & &+g_{1}\textrm{Tr}[\bar{\psi}_{6}\slashed u\gamma_{5}\psi_{6}]+g_{2}\textrm{Tr}[(\bar{\psi}_{6}\slashed u\gamma_{5}\psi_{{\bar{3}}})+\textrm{H.c.}]+g_{3}\textrm{Tr}[(\bar{\psi}_{6^{*}}^{\mu}u_{\mu}\psi_{6})+\textrm{H.c.}] \nonumber\\
 & &+g_{4}\textrm{Tr}[(\bar{\psi}_{6^{*}}^{\mu}u_{\mu}\psi_{\bar{3}})+\textrm{H.c.}]+g_{5}\textrm{Tr}[\bar{\psi}_{6^{*}}^{\nu}\slashed u\gamma_{5}\psi_{6^{*}\nu}]+g_{6}\textrm{Tr}[\bar{\psi}_{\bar{3}}\slashed u\gamma_{5}\psi_{\bar{3}}],
\end{eqnarray}
with
\begin{eqnarray}
&& D_{\mu}\psi =\partial_{\mu}\psi+\Gamma_{\mu}\psi+\psi\Gamma_{\mu}^{T},\\
 && \Gamma_{\mu}=\frac{1}{2}[\xi^{\dagger},\partial_{\mu}\xi]+\frac{i}{2}eA_{\mu}(\xi^{\dagger}Q_{B}\xi+\xi Q_{B}\xi^{\dagger}),\\
 && u_{\mu}=\frac{i}{2}\{\xi^{\dagger}, \partial_{\mu}\xi\}-\frac{1}{2}eA_{\mu}(\xi^{\dagger}Q_{B}\xi-\xi Q_{B}\xi^{\dagger}),
\end{eqnarray}
where $Q_B$ is the charge matrix of the heavy baryon. It is related to the charge matrix of the heavy quark $ \tilde Q$ and that of the light quark $Q$ through the relation $Q_B=Q+\frac{1}{2}\tilde Q$. For the charmed baryons, one has $\tilde Q_c=\text{diag}(\frac{2}{3},\frac{2}{3},\frac{2}{3})$ and $Q_{B}=\text{diag}(1,0,0)$, respectively.
$M_{\bar 3, 6,6^*}$ denote the average masses of the antitriplet, spin-$1\over 2$ sextet, spin-$3\over 2$ sextet states, respectively.

In HBChPT, the nonrelativistic form of the $\mathcal{L}_{B\phi}^{(1)} $ reads~\cite{Yan:1992gz, Jiang:2014ena}
\begin{eqnarray}
\mathcal{L}_{B\phi}^{(1)}&=&\frac{1}{2}\text{Tr}(\bar{B}_{\bar{3}}iv\cdot DB_{\bar{3}})+\text{Tr}[\bar{B}_{6}(iv\cdot D-\delta_{1})B_{6}]-\text{Tr}[\bar{B}_{6}^{*\mu}(iv\cdot D-\delta_{2})B_{6\mu}^{*}]\nonumber\\
&& +2g_{1}\text{Tr}(\bar{B}_{6}S\cdot uB_{6})+2g_{2}\text{Tr}(\bar{B}_{6}S\cdot uB_{\bar{3}}+{\rm H.c.})+g_{3}\text{Tr}(\bar{B}_{6\mu}^{*}u^{\mu}B_{6}+{\rm H.c.})\nonumber\\
 && +g_{4}\text{Tr}(\bar{B}_{6\mu}^{*}u^{\mu}B_{\bar{3}}+{\rm H.c.})+2g_{5}\text{Tr}(\bar{B}_{6}^{*\mu}S\cdot uB_{6\mu}^{*})+2g_{6}\text{Tr}(\bar{B}_{\bar{3}}S\cdot uB_{\bar{3}}),
\end{eqnarray}
where the mass differences are $\delta_1=M_{6}-M_{\bar{3}}$, $\delta_2=M_{{6^*}}-M_{\bar{3}}$ and $\delta_3=M_{{6^*}}-M_{6}$.

The $\mathcal{O}(p^2)$ Lagrangian $\mathcal{L}_{B\gamma}^{(2)}$ contributes to the leading-order magnetic moment at the tree level,
\begin{eqnarray}\label{trp2}
\mathcal{L}_{B\gamma}^{(2)} &=&-\frac{id_{2}}{8M_{N}}\text{Tr}(\bar{B}_{\bar{3}}[S^{\mu},S^{\nu}]\tilde{f}_{\mu\nu}^{+}B_{\bar{3}})-\frac{id_{3}}{8M_{N}}\text{Tr}(\bar{B}_{\bar{3}}[S^{\mu},S^{\nu}]B_{\bar{3}})\text{Tr}(f_{\mu\nu}^{+})\nonumber\\
 && -\frac{id_{5}}{4M_{N}}\text{Tr}(\bar{B_{6}}[S^{\mu},S^{\nu}]\tilde{f}_{\mu\nu}^{+}B_{6})-\frac{id_{6}}{4M_{N}}\text{Tr}(\bar{B_{6}}[S^{\mu},S^{\nu}]B_{6})\text{Tr}(f_{\mu\nu}^{+})\nonumber\\
 && -\frac{id_{8}}{2M_{N}}\text{Tr}(\bar{B}_{6^{*}}^{\mu}\tilde{f}_{\mu\nu}^{+}B_{6^{*}}^{\nu})-\frac{id_{9}}{2M_{N}}\text{Tr}(\bar{B}_{6^{*}}^{\mu}B_{6^{*}}^{\nu})\text{Tr}(f_{\mu\nu}^{+})\nonumber\\
 && -\frac{2if_{2}}{M_{N}}\text{Tr}(\bar{B}_{\bar{3}}\tilde{f}_{\mu\nu}^{+}[S^{\mu},S^{\nu}]B_{6})-i\frac{f_{4}}{2M_{N}}\text{Tr}(\bar B_{6}^{*\mu}\tilde{f}_{\mu\nu}^{+}S^{\nu}{B}_{\bar{3}})\nonumber\\
 && -i\frac{f_{3}}{2M_{N}}\text{Tr}(\bar{B}_{6}^{*\mu}\tilde{f}_{\mu\nu}^{+}S^{\nu}B_{6})-i\frac{\tilde{f_{3}}}{2M_{N}}\text{Tr}(\bar{B}_{6}^{*\mu}S^{\nu}B_{6})\text{Tr}(f_{\mu\nu}^{+})+\textrm{H.c.}.
 \end{eqnarray}
 where $M_N$ is the nucleon mass. The tensor fields $\tilde{f}_{\mu\nu}^{+}$ and $\text{Tr}(f_{\mu\nu}^{+})$ are defined as
 \begin{eqnarray}
 && f_{\mu\nu}^{R}=f_{\mu\nu}^{L}=-eQ_{B}(\partial_{\mu}A_{\nu}-\partial_{\nu}A_{\mu}),\\
 && f_{\mu\nu}^{\pm}=\xi^{\dagger}f_{\mu\nu}^{R}\xi\pm\xi^{\dagger}f_{\mu\nu}^{L}\xi,\\
 && \tilde{f}_{\mu\nu}^{\pm}=f_{\mu\nu}^{\pm}-\frac{1}{3}\text{Tr}(f_{\mu\nu}^{\pm}).
\end{eqnarray}
 Since $\text{Tr}(Q)=0$ and
$\text{Tr}(\tilde Q)\neq0$, the $\tilde{f}_{\mu\nu}^{+}$ and $\text{Tr}(f_{\mu\nu}^{+})$
represent the contributions from the light and heavy quarks,
respectively. The two building blocks are in the octet and singlet flavor representations, respectively. In the flavor space, $3\otimes\bar{3}=1\oplus8$ and $6\otimes\bar{6}=1\oplus8\oplus27$. The $d_{2,5,8}$ and $d_{3,6,9}$ terms
correspond to the $8\otimes8\rightarrow1$ and $1\otimes1\rightarrow 1$, respectively. In the antitriplet, the $J^P$ of the light diquark is $0^+$. The coupling constant $d_2$ vanishes since the M1 transition $|0^+\rangle\rightarrow |0^+\rangle\gamma$ is forbidden. For the $B_{6}/B^{\mu}_{6^*}\rightarrow B_{\bar 3}\gamma $ transition, the heavy baryons form the $3\otimes 6=8\oplus 10$ flavor representation. Thus, they can only couple with $\tilde{f}_{\mu\nu}^{+}$ to form the flavor singlet. The leading-order $B_{6}/B^{\mu}_{6^*}\rightarrow B_{\bar 3}\gamma $ transition totally arises from the dynamics of the light quark sector.

The $\mathcal O(p^2)$ Lagrangians constructed from other building blocks do not contribute to the $\mathcal O(p^2)$ radiative decays. For instance, the following $\chi_{\pm}$ is  $\mathcal O(p^2)$,
 \begin{eqnarray}
 & \chi_{\pm}=\xi^{\dagger}\chi\xi^{\dagger}\pm\xi\chi^{\dagger}\xi,\\
 & \chi=2B_{0}\text{diag}(m_{u},m_{d},m_{s}),
\end{eqnarray}
where $B_{0}=-\frac{1}{3F_{\phi}^{2}}\langle\bar{q}q\rangle$ is a parameter related to the quark condensate, $m_{u,d,s}$ denotes the current quark mass. At the leading order, the $m_{u,d}$ is ignored and $2B_0 m_s$ are absorbed into the coupling constant. We obtain $\chi_+=\text{diag}(0,0,1)$. We can construct the $\mathcal O(p^2)$ Lagrangians with $\chi_{+}$, for instance, $\text{Tr}(\bar B_{\bar 3}\chi_+B_6)$. However, they do not contribute to the radiative decay amplitude at $\mathcal O(p^2)$.

The $\mathcal{O}(p^2)$ $B\phi\phi$ vertex arising from $\mathcal{L}_{B\phi\phi}^{(2)}$ contributes to the $\mathcal O(p^4)$ decay amplitude,
 \begin{eqnarray}\label{diph}
\mathcal{L}_{B\phi\phi}^{(2)} &=&\frac{a_{1}}{M_{N}}\text{Tr}(\bar{B}_{\bar{3}}[S^{\mu},S^{\nu}][u_{\mu},u_{\nu}]B_{6})+\frac{a_{2}}{M_{N}}\text{Tr}(\bar{B}_{\bar{3}ab}[S^{\mu},S^{\nu}]u_{i\mu}^{b}u_{j\nu}^{a}B_{6}^{ij})\nonumber\\
 && +\frac{a_{3}}{M_{N}}\text{Tr}(\bar{B}_{\bar{3}}S^{\mu}[u_{\mu},u_{\nu}]B_{6^{*}}^{\nu})+\frac{a_{4}}{M_{N}}\text{Tr}(\bar{B}_{\bar{3}ab}S^{\mu}u_{i\mu}^{b}u_{j\nu}^{a}B_{6^{*}}^{\nu ij})\nonumber\\
 && +\frac{a_{5}}{M_{N}}\text{Tr}(\bar{B}_{6}S^{\mu}[u_{\mu},u_{\nu}]B_{6^{*}}^{\nu})+\textrm{H.c.}.
\end{eqnarray}
The $u_{\mu}$ and $u_{\nu}$ form the flavor representations $8\otimes8=1\oplus8_{1}\oplus8_{2}\oplus10\oplus\overline {10}\oplus27$
as illustrated in Table~\ref{fig1}. For the transitions from the sextet to the antitriplet, the baryon building blocks have $6\otimes3=8\oplus10$. The $a_{1,3}$ and $a_{2,4}$ terms
correspond to the $8\otimes8_{1}\rightarrow1$ and $10\otimes\overline{10}\rightarrow1$,
respectively. The term $\text{Tr}(\bar{B}_{\bar{3}}[S^{\mu},S^{\nu}]\{u_{\mu},u_{\nu}\}B_{6})$ corresponding to $8\otimes8_{2}\rightarrow1$, vanishes due to the
antisymmetry of the Lorentz indices $\mu$ and $\nu$. For the transition $B^\mu_{6^*}\rightarrow B_6\gamma$, the baryons form the $6\otimes \bar 6=1\oplus 8 \oplus10$ flavor representations. There should have existed four independent interaction terms corresponding to $8\otimes8_{1}\rightarrow1$, $8\otimes8_{2}\rightarrow1$, $1\otimes1 \rightarrow1$ and $27\otimes 27\rightarrow1$. The explicit forms of the Lagrangians are $\text{Tr}(\bar{B}_{6}S^{\mu}[u_{\mu},u_{\nu}]B_{6^{*}}^{\nu}) $, $\text{Tr}(\bar{B}_{6}S^{\mu}\{u_{\mu},u_{\nu}\}B_{6}^{*\nu})$, $\text{Tr}(\bar{B}_{6}S^{\mu}B_{6}^{*\nu})\text{Tr}(u_{\mu},u_{\nu})$ and $\text{Tr}(\bar{B}_{6ab}S^{\mu}\{u_{i\mu}^{a}u_{j\nu}^{b}\}_{\{ij\}}^{\{ab\}}B_{6^{*}}^{\nu ij})$, respectively. In diagram (d) in Fig.~\ref{allloop}, the vertices $B\phi\phi$ arising from the last three Lagrangian terms are symmetric for the opposite charged pseudoscalar mesons, while the $\phi\phi\gamma $ vertex is antisymmetric. Then the loops with the opposite charged intermediate pseudoscalar mesons cancel out.  Therefore, the above three Lagrangian terms do not contribute to the radiative decay.

\begin{table}
\caption{The $u_{\mu}$ $\otimes$ $u_{\nu}$ may be in the $8\otimes8=1\oplus8_{1}\oplus8_{2}\oplus10\oplus\overline{10}\oplus27$ flavor representations. $\{\}_{\{bj\}}^{\{ai\}}$ represents that the scripts $b(a)$ and the $j(i)$ are symmetrized. $\mathcal{S}_{123}$ is the symmetrization operator for the subscripts $1$, $2$, and $3$. $\epsilon_{ijk}$ is the Levi-Civita symbol.}
\label{fig1}
\begin{tabular}{c|c|c|c|c|c}
\toprule[1pt]\toprule[1pt]
1 &$ 8_{1}$ & $8_{2}$& $10$ & $\overline{10}$ & 27 \tabularnewline
\midrule[1pt]
$\text{Tr}(u_{\mu}u_{\nu})$ &$[u_{\mu},u_{\nu}]$ & $\left\{ u_{\mu},u_{\nu}\right\} $ &$\mathcal{S}_{ijk}\mu_{a}^{\mu i}\mu_{b}^{\nu j}\epsilon^{abk}$ & $\mathcal{S}_{abc}\mu_{a}^{\mu i}\mu_{b}^{\nu j}\epsilon_{ijc}$& $\{u_{b}^{\mu a}u_{j}^{\nu i}\}_{\{bj\}}^{\{ai\}}$ \tabularnewline
\bottomrule[1pt]\bottomrule[1pt]
\end{tabular}
\end{table}

The decay $|{1\over 2}\rangle \rightarrow |{1\over 2}\rangle \gamma$ is the M1 transition.  The transition $|\frac{3}{2}\rangle \rightarrow |\frac{1}{2}\rangle \gamma$ may happen through the M1 and E2 transitions. The Lagrangian at $\mathcal{O}(p^{3})$
contributes,
 \begin{eqnarray}
\mathcal{L}^{(3)}_{B\gamma} &=&\frac{n}{8M_{N}^{2}M_{6}}\text{Tr}(\bar{\psi}_{\bar{3}}\nabla_{\lambda}\tilde{f}_{\mu\nu}^{+}\sigma^{\mu\nu}D^{\lambda}\psi_{6})+\frac{n_{1}}{8M_{N}^{2}M_{6^{*}}}\text{Tr}(\bar{\psi}_{\bar{3}}\nabla_{\lambda}\tilde{f}_{\mu\nu}^{+}\gamma^{\lambda}\gamma_{5}iD^{\mu}\psi_{6^{*}}^{\nu})\nonumber\\
&&+\frac{n_{2}}{8M_{N}^{2}M_{6^{*}}}\text{Tr}(\bar{\psi}_{\bar{3}}\nabla_{\lambda}\tilde{f}_{\mu\nu}^{+}\gamma^{\mu}\gamma_{5}iD^{\lambda}\psi_{6^{*}}^{\nu})+\frac{m_{1}}{8M_{N}^{2}M_{6^{*}}}\text{Tr}(\bar{\psi}_{6}\nabla_{\lambda}\tilde{f}_{\mu\nu}^{+}\gamma^{\lambda}\gamma_{5}iD^{\mu}\psi_{6^{*}}^{\nu})\nonumber\\
&&+\frac{m_{2}}{8M_{N}^{2}M_{6^{*}}}\text{Tr}(\bar{\psi}_{6}\nabla_{\lambda}\tilde{f}_{\mu\nu}^{+}\gamma^{\mu}\gamma_{5}iD^{\lambda}\psi_{6^{*}}^{\nu})+\frac{\tilde{m}_{1}}{8M_{N}^{2}M_{6^{*}}}\text{Tr}[\bar{\psi}_{6}\nabla_{\lambda}\text{Tr(}f_{\mu\nu}^{+})\gamma^{\lambda}\gamma_{5}iD^{\mu}\psi_{6^{*}}^{\nu}]\nonumber\\
&&+\frac{\tilde{m}_{2}}{8M_{N}^{2}M_{6^{*}}}\text{Tr}[\bar{\psi}_{6}\text{Tr(}\nabla_{\lambda}f_{\mu\nu}^{+})\gamma^{\mu}\gamma_{5}iD^{\lambda}\psi_{6^{*}}^{\nu}]+\text{H.c.}. 
\end{eqnarray}
where $n$, $n_2$, $m_2$, and $\tilde m_2$ terms contribute to the $G_1$. They cancel the divergences of the $\mathcal O(p^3)$ loop diagrams. The finite terms have the same structures as those in the $\mathcal O(p^2)$ tree diagrams when the same meson decay constants are adopted $F_\pi=F_K=F_{\eta}$. Then they can be  absorbed  into the lower order $f_{2-4}$ and $\tilde f_3$ terms in Eq. (\ref{trp2}). The $n_1$, $m_1$ and $\tilde{m}_1$ terms contribute to $G_2$, which contributes to the lowest-order  E2 transition. 

The  nonrelativistic form of $\mathcal{L}^{(3)}_{B\gamma}$ is 
 \begin{eqnarray}
\mathcal{L}^{(3)}_{B\gamma} &=&-\frac{n}{4M_{N}^{2}}\text{Tr}(\bar{\psi}_{\bar{3}}\nabla_{\lambda}\tilde{f}_{\mu\nu}^{+}[S^{\mu},S^{\nu}]v^{\lambda}\psi_{6})+\frac{n_{1}}{4M_{N}^{2}}\text{Tr}(\bar{B}_{\bar{3}}\nabla_{\lambda}\tilde{f}_{\mu\nu}^{+}S^{\lambda}v^{\mu}B_{6^{*}}^{\nu})+\frac{n_{2}}{4M_{N}^{2}}\text{Tr}(\bar{B}_{\bar{3}}\nabla_{\lambda}\tilde{f}_{\mu\nu}^{+}S^{\mu}v^{\lambda}B_{6^{*}}^{\nu})\nonumber\\
&&+\frac{m_{1}}{4M_{N}^{2}}\text{Tr}(\bar{B}_6\nabla_{\lambda}\tilde{f}_{\mu\nu}^{+}S^{\lambda}v^{\mu}B_{6^{*}}^{\nu})+\frac{m_{2}}{4M_{N}^{2}}\text{Tr}(\bar{B}_6\nabla_{\lambda}\tilde{f}_{\mu\nu}^{+}S^{\mu}v^{\lambda}B_{6^{*}}^{\nu})\nonumber\\
&&+\frac{\tilde{m}_{1}}{4M_{N}^{2}}\text{Tr}[\bar{B}_{6}\nabla_{\lambda}\text{Tr(}f_{\mu\nu}^{+})S^{\lambda}v^{\mu}B_{6^{*}}^{\nu}]+\frac{\tilde{m}_{2}}{4M_{N}^{2}}\text{Tr}[\bar{B}_{6}\text{Tr(}\nabla_{\lambda}f_{\mu\nu}^{+})S^{\mu}v^{\lambda}B_{6^{*}}^{\nu}]+\text{H.c.} .
\end{eqnarray}

At $\mathcal{O}(p^{4})$, the Lagrangian that contributes at the tree level is,
 \begin{eqnarray}
\mathcal{L}_{B\gamma}^{(4)}&=&\frac{c_{1}}{8m_{N}}\text{Tr}(\bar{\psi}_{\bar{3}}\chi_{+}\sigma_{\mu\nu}\psi_{6})\text{Tr}(f_{\mu\nu}^{+})+\frac{c_{2}}{8m_{N}}\text{Tr}(\bar{\psi}_{\bar{3}}\{ \chi_{+},\tilde {f}_{\mu\nu}^{+}\}\sigma_{\mu\nu}\psi_{6}) +\frac{c_{3}}{8m_{N}}\text{Tr}(\bar{\psi}_{\bar{3}}^{ab}\chi_{a+}^{i}\tilde{f}_{b\mu\nu}^{j+}\sigma_{\mu\nu}\psi_{6ij})\nonumber\\
 && -\frac{ih_{1}}{8m_{N}}\text{Tr}(\bar{\psi}_{\bar{3}}\chi_{+}\gamma_{\mu}\gamma_{5}\psi_{6^{*}}^{\nu})\text{Tr}(f_{\mu\nu}^{+})
 -\frac{ih_{2}}{8m_{N}}\text{Tr}(\bar{\psi}_{\bar{3}}\{\chi_{+},\tilde {f}_{\mu\nu}^{+}\}\gamma_{\mu}\gamma_{5}\psi_{6^{*}}^{\nu})-\frac{ih_{3}}{8m_{N}}\text{Tr}(\bar{\psi}_{\bar{3}}^{ab}\chi_{a+}^{i}\tilde {f}_{b\mu\nu}^{j+}\gamma_{\mu}\gamma_{5}\psi_{6^{*}ij}^{\nu})\nonumber\\
 &&-\frac{il_{1}}{8m_{N}}\text{Tr}(\bar{\psi}_{6}\chi_{+}\gamma_{\mu}\gamma_{5}\psi_{6^{*}}^{\nu})\text{Tr}(f_{\mu\nu}^{+}) -\frac{il_{2}}{8m_{N}}\text{Tr}(\bar{\psi}_{6}^{ab}\{\chi_{+}\tilde {f}_{\mu\nu}^{+}\}_{ab}^{ij}\gamma_{\mu}\gamma_{5}\psi_{6^{*}ij}^{\nu})+\text{H.c.}.
\end{eqnarray}
As illustrated in Tables~\ref{p4lagrangian1} and~\ref{p4lagrangian2}, there are five and six independent Lagrangian terms for the transitions $B_{6}/B^{\mu}_{6^*} \rightarrow B_{\bar 3}\gamma$ and $B^{\mu}_{6^*}\rightarrow B_ 6\gamma$, respectively. The leading order expansion of the operator $[\chi_+,{\tilde f}_{\mu\nu}]$ vanishes, since they are diagonal matrices. Many terms are absorbed by the other Lagrangians.

In the nonrelativistic limit, $\mathcal{L}_{B\gamma}^{(4)}$ is written as
 \begin{eqnarray}\label{trp4}
\mathcal{L}_{B\gamma}^{(4)} & =&\frac{-ic_{1}}{4m_{N}}\text{Tr}(\bar{B}_{\bar{3}}\chi_{+}[S_{\mu,}S_{\nu}]B_{6})\text{Tr}(f_{\mu\nu}^{+})+\frac{-ic_{2}}{4m_{N}}\text{Tr}(\bar{B}_{{\bar{3}}}\{ \chi_{+},\tilde {f}_{\mu\nu}^{+}\} [S_{\mu,}S_{\nu}]B_{6})+\frac{-ic_{3}}{4m_{N}}\text{Tr}(\bar{B}_{\bar{3}}^{ab}\chi_{a+}^{i}\tilde {f}_{b\mu\nu}^{j+}[S_{\mu,}S_{\nu}]B_{6ij})\nonumber\\
 && +\frac{-ih_{1}}{4m_{N}}\text{Tr}(\bar{B}_{\bar{3}}\chi_{+}S_{\mu}B_{6^{*}}^{\nu})\text{Tr}(f_{\mu\nu}^{+})+\frac{-ih_{2}}{4m_{N}}\text{Tr}(\bar{B}_{{\bar{3}}}\{\chi_{+},\tilde {f}_{\mu\nu}^{+}\}S_{\mu}B_{6^{*}}^{\nu})+\frac{-ih_{3}}{4m_{N}}\text{Tr}(\bar{B}_{\bar{3}}^{ab}\chi_{a+}^{i}\tilde {f}_{b\mu\nu}^{j+}S_{\mu}B_{6^{*}ij}^{\nu})\nonumber\\
 &&+\frac{-il_{1}}{4m_{N}}\text{Tr}(\bar{B}_{6}\chi_{+}S_{\mu}B_{6^{*}}^{\nu})\text{Tr}(f_{\mu\nu}^{+})+\frac{-il_{2}}{4m_{N}}\text{Tr}(\bar{B}_{6}^{ab}\{\chi_{+},\tilde {f}_{\mu\nu}^{+}\}_{ab}^{ij}S_{\mu}B_{6^{*}ij}^{\nu})+\text{H.c.}.
\end{eqnarray}

\begin{table} 
\caption{The possible flavor structures constructed by two baryons in the $\mathcal{O}(p^{4})$
Lagrangians. }\label{p4lagrangian1}
\begin{tabular}{c|c|c|c|c|c}
\toprule[1pt]\toprule[1pt] 
Group representation & $3\otimes6\rightarrow8$ & $3\otimes6\rightarrow10$ & $6\otimes6\rightarrow1$ & $6\otimes6\rightarrow8$ & $6\otimes6\rightarrow27$\tabularnewline
\hline 
Flavor structure & $\bar{B}_{\bar{3}}^{ab}B_{6ca}$ & $\bar{B}_{\bar{3}}^{ab}B_{6ij}$ & $\text{Tr}(\bar{B}_{6}B_{6})$ & $\bar{B}_{6}^{ab}B_{6ca}$ & $\bar{B}_{6}^{ab}B_{6ij}$\tabularnewline
\bottomrule[1pt]\bottomrule[1pt]
\end{tabular}
\end{table}

\begin{table}
\caption{The possible flavor structures constructed by $\chi_+$, $\tilde f_{\mu\nu}$ or $\text{Tr}(f_{\mu\nu})$ for the $\mathcal{O}(p^{4})$
Lagrangians. These structures combine with those 
in Table~\ref{p4lagrangian1} to form the Lagrangians according to group representations:
$\bar{8}\otimes8\rightarrow1$, $\overline{10}\otimes10\rightarrow1$,
$27\otimes27\rightarrow1$ and $1\otimes1\rightarrow1$. The three
$\{~\}$ in the third or sixth rows correspond to $\mathcal{L}_{36}$, $\mathcal{L}_{36^*}$ and $\mathcal{L}_{66^*}$, respectively. The ``$\text{ab.}f_{i}$''
means that the LEC can be absorbed by $f_{i}$. And $\{-\}$ represents that the corresponding group representation does not exist. }\label{p4lagrangian2}
\begin{tabular}{c|c|c|c|c}
\toprule[1pt]\toprule[1pt] 
Group representation & $\ensuremath{1\otimes1\rightarrow1}$ & $\ensuremath{1\otimes8\rightarrow8}$ & $\ensuremath{8\otimes1\rightarrow8}$ & $\ensuremath{8\times8\rightarrow1}$\tabularnewline
\hline 
Flavor structure & $\text{Tr}(\chi_{+})\text{Tr}(f_{\mu\nu}^{+})$ & $\text{Tr}(\chi_{+})\tilde{f}_{\mu\nu}^{+}$ & $\chi_{+}\text{Tr}(\tilde{f}_{\mu\nu}^{+})$ & $\text{Tr}(\chi_{+}\tilde{f}_{\mu\nu}^{+})$\tabularnewline
\hline 
LECs & $\{-\}\{-\}\{\text{ab.}\tilde{f_{3}}\}$ & $\{\text{ab.}f_{2}\}\{\text{ab.}f_{3}\}\{\text{ab.}f_{4}\}$ & $\{c_{1}\}\{h_{1}\}\{l_{1}\}$ & $\{-\}\{-\}\{\text{ab.}\tilde{f_{3}}\}$\tabularnewline
\midrule[1pt]
Group representation & $8\otimes8\rightarrow8_{1}$ & $\ensuremath{8\otimes8\rightarrow8_{2}}$ & $\ensuremath{8\otimes8\rightarrow27}$ & $\ensuremath{8\otimes8\rightarrow\bar{10}}$\tabularnewline
\hline 
Flavor structure & $[\chi_{+},\tilde{f}_{\mu\nu}^{+}]$ & $\{\chi_{+},\tilde{f}_{\mu\nu}^{+}\}$ & $\{\chi_{+},\tilde{f}_{\mu\nu}^{+}\}_{ab}^{ij}$ & $(\chi_{+})_{a}^{i}(\tilde{f}_{\mu\nu}^{+})_{b}^{j}$\tabularnewline
\hline 
LECs & vanishing & $\{c_{2}\}\{h_{2}\}\{\text{ab.}l_{1}\}$ & $\{-\}\{-\}\{l_{2}\}$ & $\{c_{3}\}\{h_{3}\}\{-\}$\tabularnewline
\bottomrule[1pt]\bottomrule[1pt]
\end{tabular}
\end{table}

\section{analytical expression}\label{sec3}

\subsection{$B_6(p')\rightarrow B_{\bar 3}(p)+\gamma(q)$}

 At the leading order, the $\mathcal O(p^2)$ tree diagram in Fig.~\ref{tree} stems from the $\mathcal L^{(2)}_{B\gamma}$ and contributes to $\mathcal O(p^2)$ decay amplitude and $\mathcal O(p)$ transition magnetic moment,
 \begin{eqnarray}
&\mu_{\text{tree}}^{(1)}(\Sigma^+_c\rightarrow \Lambda^+_c\gamma)=4\sqrt{2}f_{2},\quad\mu_{\text{tree}}^{(1)}(\Xi_c^{'+}\rightarrow\Xi_c^{+}\gamma)=4\sqrt{2}f_{2},\quad\mu_{\text{tree}}^{(1)}(\Xi_c^{'0}\rightarrow\Xi_c^{0}\gamma)=0.
\end{eqnarray}
The superscript denotes the chiral order. The transition magnetic moment is in the unit of nuclear magneton.

At NLO, the results from the tree diagrams are 
\begin{eqnarray}
&\mu_{\text{tree}}^{(2)}(\Sigma^+_c\rightarrow \Lambda^+_c\gamma)=\frac{n \ell}{\sqrt 2 M_N},\quad\mu_{\text{tree}}^{(2)}(\Xi_c^{'+}\rightarrow\Xi_c^{+}\gamma)=\frac{n \ell}{\sqrt 2 M_N},\quad\mu_{\text{tree}}^{(2)}(\Xi_c^{'0}\rightarrow\Xi_c^{0}\gamma)=0.
\end{eqnarray}
with $~\ell=v\cdot q$.
At NLO, the chiral corrections come from the loop diagrams (a), (b) and (i)-(l). After the integration, the amplitudes of the diagrams (i)-(l) vanish due to $S\cdot v=0$. The (a) and (b) diagrams contribute to the $\mathcal O(p^3)$ decay amplitude and $\mathcal O(p^2)$ transition magnetic moment:
\begin{equation}
\mu^{(2)}=C_{\phi}^{(3)}M_{N}\left[\frac{g_{3}g_{4}}{4F_{\phi}^{2}}\frac{4}{d-1}n_{1}^{\text{II}}(-\delta_{3},-\delta_{3}-\ell)-\frac{g_{1}g_{2}}{F_{\phi}^{2}}n_{1}^{\text{II}}(0,-\ell)\right],~~ \ell=v \cdot q
\end{equation}
where $n_{1}^{\text{II}}$ is the finite part of the loop integral and its explicit form is given in the Appendix. $d$ is the dimension. $\phi$ represents the intermediate pseudoscalar meson in the loop. 
 $C_{\phi}^{(3)}$ is the coefficient for the loops as illustrated in Table~\ref{cg63}.

At NNLO, the chiral corrections come from the $\mathcal O(p^4)$ tree diagram in Fig.~\ref{tree} and the loop diagrams (c)-(h) and (m)-(p) in Fig.~\ref{allloop}. The $\mathcal O(p^3)$ magnetic moments from the tree diagram read,
\begin{eqnarray}
 \mu_{\text{tree}}^{(3)}(\Sigma_{c}^{+}\rightarrow\Lambda_{c}^{+})=0,\,\,\,\mu_{\text{tree}}^{(3)}(\Xi_{c}^{'+}\rightarrow\Xi_{c}^{+})=\frac{2\left(c_{2}+c_{3}\right)-3c_{1}}{3\sqrt{2}},\,\,\,\mu_{\text{tree}}^{(3)}(\Xi_{c}^{'0}\rightarrow\Xi_{c}^{0})=-\frac{3c_{1}-2c_{2}+c_{3}}{3\sqrt{2}}. \nonumber \\
\end{eqnarray}
 The $\mathcal O(p^3)$ magnetic moments from the loop diagrams read

%\begin{eqnarray}
%\mu_{l}^{(c-d)}&=&\delta^{\phi}\frac{f_{2}}{F_{\phi}^{2}}\left(\frac{m_{\phi}^{2}}{16\pi^{2}}\ln\frac{m_{\phi}^{2}}{\lambda^{2}}\right)+\gamma_{1}^{\phi}\frac{a_{1}}{F_{\phi}^{2}}\left(\frac{m_{\phi}^{2}}{32\pi^{2}}\ln\frac{m_{\phi}^{2}}{\lambda^{2}}\right)+\gamma_{2}^{\phi}\frac{a_{2}}{2F_{\phi}^{2}}\left(\frac{m_{\phi}^{2}}{32\pi^{2}}\ln\frac{m_{\phi}^{2}}{\lambda^{2}}\right),\\
% \mu_{l}^{(e-h)}&=&\alpha_{6}^{\phi}\frac{g_{1}g_{2}}{F_{\phi}^{2}}\frac{d-3}{4}\Lambda_2(0,-\ell)+\alpha_{6^{*}}^{\phi}\frac{g_{3}g_{4}}{4F_{\phi}^{2}}\left(\frac{8}{1-d}+\frac{4(5-d)}{(d-1)^{2}}\right)\Lambda_2(-\ell-\delta_{3},-\delta_{3})\nonumber \\
% &~~&-\alpha_{66^{*}}^{\phi}\frac{g_{2}g_{3}}{F_{\phi}^{2}}\frac{3-d}{d-1}\Lambda_2(-\ell, -\delta_{3})-\alpha_{66^{*}}^{\phi}\frac{g_{1}g_{4}}{F_{\phi}^{2}}\frac{3-d}{d-1}\Lambda_2(-\delta_{3}-\ell,0)\nonumber \\
% &~~& +\beta^{\phi}\frac{8f_{2}g_{2}^{2}}{F_{\phi}^{2}}\frac{d-3}{4}\Lambda_2(-\ell,\delta_1)-\beta^{\phi}\frac{g_{2}g_{4}f_{4}}{F_{\phi}^{2}}\frac{3-d}{d-1}\Lambda_2(-\ell-\delta_{3},\delta_{1}),\\
% \mu_{l}^{(m-p)} &=&\frac{1}{2}\Big[O^{\phi}\frac{g_{1}^{2}}{F_{\phi}}\frac{1-d}{4}J_{2}'(0)+N^{\phi}\frac{g_{2}^{2}}{F_{\phi}^{2}}\frac{1-d}{4}J_{2}'(\delta_{1})+O^{\phi}\frac{g_{3}^{2}}{4F_{\phi}^{2}}(2-d)J_{2}'(-\delta_{3})\nonumber \\
% &~~& +N_{3}^{\phi}\frac{g_{2}^{2}}{F_{\phi}^{2}}\frac{1-d}{4}J_{2}'(-\ell)+N_{3}^{\phi}\frac{g_{4}^{2}}{4F_{\phi}^{2}}(2-d)J_{2}'(-\ell-\delta_{3})\Big]\mu_{\text{tree}}^{(2)},
% \end{eqnarray}
\begin{eqnarray}
\mu_{l}^{c}&=&\delta^{\phi}\frac{f_{2}}{F_{\phi}^{2}}\left(\frac{m_{\phi}^{2}}{16\pi^{2}}\ln\frac{m_{\phi}^{2}}{\lambda^{2}}\right),\\
\mu_{l}^{d}&=&\gamma_{1}^{\phi}\frac{a_{1}}{F_{\phi}^{2}}\left(\frac{m_{\phi}^{2}}{32\pi^{2}}\ln\frac{m_{\phi}^{2}}{\lambda^{2}}\right)+\gamma_{2}^{\phi}\frac{a_{2}}{2F_{\phi}^{2}}\left(\frac{m_{\phi}^{2}}{32\pi^{2}}\ln\frac{m_{\phi}^{2}}{\lambda^{2}}\right),\\
\mu_{l}^{(e)}&=&\alpha_{6^{*}}^{\phi}\frac{g_{3}g_{4}}{4F_{\phi}^{2}}\left(\frac{8}{1-d}+\frac{4(5-d)}{(d-1)^{2}}\right)\Lambda_{2}(-\ell-\delta_{3},-\delta_{3}),\\
\mu_{l}^{(f)}&=&\alpha_{6}^{\phi}\frac{g_{1}g_{2}}{F_{\phi}^{2}}\frac{d-3}{4}\Lambda_{2}(0,-\ell)+\beta^{\phi}\frac{8f_{2}g_{2}^{2}}{F_{\phi}^{2}}\frac{d-3}{4}\Lambda_{2}(-\ell,\delta_{1}),\\
\mu_{l}^{(g)}&=&-\alpha_{66^{*}}^{\phi}\frac{g_{2}g_{3}}{F_{\phi}^{2}}\frac{3-d}{d-1}\Lambda_{2}(-\ell,-\delta_{3}),\\
\mu_{l}^{(h)}&=&-\alpha_{66^{*}}^{\phi}\frac{g_{1}g_{4}}{F_{\phi}^{2}}\frac{3-d}{d-1}\Lambda_{2}(-\delta_{3}-\ell,0)-\beta^{\phi}\frac{g_{2}g_{4}f_{4}}{F_{\phi}^{2}}\frac{3-d}{d-1}\Lambda_{2}(-\ell-\delta_{3},\delta_{1}).\\
\mu_{l}^{(m)}&=&\frac{1}{2}N_{3}^{\phi}\frac{g_{4}^{2}}{4F_{\phi}^{2}}(2-d)J_{2}'(-\ell-\delta_{3})\mu_{\text{tree}}^{(2)},\\
\mu_{l}^{(n)}&=&\frac{1}{2}N_{3}^{\phi}\frac{g_{2}^{2}}{F_{\phi}^{2}}\frac{1-d}{4}J_{2}'(-\ell),\\
\mu_{l}^{(o)}&=&\frac{1}{2}O^{\phi}\frac{g_{3}^{2}}{4F_{\phi}^{2}}(2-d)J_{2}'(-\delta_{3})\mu_{\text{tree}}^{(2)},\\
\mu_{l}^{(p)}&=&\frac{1}{2}\Big[O^{\phi}\frac{g_{1}^{2}}{F_{\phi}}\frac{1-d}{4}J_{2}'(0)+N^{\phi}\frac{g_{2}^{2}}{F_{\phi}^{2}}\frac{1-d}{4}J_{2}'(\delta_{1})\mu_{\text{tree}}^{(2)}\Big]\mu_{\text{tree}}^{(2)}.
 \end{eqnarray}
 where $J_2$, $J'_2$ and $\Lambda_2 $ are the finite parts of the loop integrals. $\delta^{\phi}$, $\gamma^{\phi}_{1,2}$, $a^{\phi}_{66^*}$ and other coefficients for the loops are listed in Table~\ref{cg63}.

\begin{table}[!htp]
 \caption{The coefficients for the tree and loop diagrams that contribute to the amplitudes in the radiative decays $B_6/B^{\mu}_{6^*}\rightarrow B_{\bar 3}\gamma$ up to NNLO. }\label{cg63}
\begin{tabular}{c|c|ccc}
\toprule[1pt]\toprule[1pt]
\multicolumn{2}{c|}{$B_{6}\rightarrow B_{\bar{3}}\gamma$} & $\Sigma_{c}^{+}\rightarrow\Lambda_{c}$ & $\Xi_{c}^{'+}\rightarrow\Xi_{c}^{+}$ & $\Xi_{c}^{'0}\rightarrow\Xi_{c}^{0}$\tabularnewline
\hline 
\multicolumn{2}{c|}{$B_{6^{*}}^{\mu}\rightarrow B_{\bar{3}}\gamma$} & $\Sigma_{c}^{*+}\rightarrow\Lambda_{c}$ & $\Xi_{c}^{*+}\rightarrow\Xi_{c}^{+}$ & $\Xi_{c}^{*0}\rightarrow\Xi_{c}^{0}$\tabularnewline
\midrule[1pt]
$\mathcal{O}(p^{2})$ Tree & $C^{(2)}$ & $\frac{1}{\sqrt{2}}$ & $\frac{1}{\sqrt{2}}$ & $0$\tabularnewline
\midrule[1pt]
\multirow{2}{*}{$(a),(b)$} & $C_{\pi}^{(3)}$ & $2\sqrt{2}$ & $\frac{1}{\sqrt{2}}$ & $-\frac{1}{\sqrt{2}}$\tabularnewline
 & $C_{k}^{(3)}$ & $\frac{1}{\sqrt{2}}$ & $2\sqrt{2}$ & $\frac{1}{\sqrt{2}}$\tabularnewline
\midrule[1pt]
\multirow{2}{*}{$(c)$} & $\delta{}^{\pi}$ & $-4\sqrt{2}$ & $-2\sqrt{2}$ & $2\sqrt{2}$\tabularnewline

 & $\delta{}^{k}$ & $-2\sqrt{2}$ & $-4\sqrt{2}$ & $-2\sqrt{2}$\tabularnewline
\midrule[1pt]
\multirow{4}{*}{$(d)$} & $\gamma^{\pi}$ & $2\sqrt{2}$ & $\sqrt{2}$ & $-\sqrt{2}$\tabularnewline
 & $\gamma^{k}$ & $\sqrt{2}$ & $2\sqrt{2}$ & $\sqrt{2}$\tabularnewline
 & $\gamma_{2}^{\pi}$ & $\sqrt{2}$ & $0$ & $0$\tabularnewline
 & $\gamma_{2}^{k}$ & $0$ & $\sqrt{2}$ & $0$\tabularnewline
\midrule[1pt]

\multirow{8}{*}{$(e),(f),(g),(h)$} & $\alpha_{6}^{\pi}$ & $-\sqrt{2}d_{5}$ & $-\frac{d_{5}-6d_{6}}{4\sqrt{2}}$ & $\frac{3d_{6}}{2\sqrt{2}}$\tabularnewline
 & $\alpha_{6}^{k}$ & $-\frac{d_{5}}{2\sqrt{2}}$ & $-\frac{1}{\sqrt{2}}(\frac{13}{6}d_{5}+d_{6})$ & $-\frac{1}{\sqrt{2}}(\frac{1}{6}d_{5}+d_{6})$\tabularnewline
 & $\alpha_{6}^{\eta}$ & $0$ & $-\frac{d_{5}+6d_{6}}{12\sqrt{2}}$ & $\frac{d_{5}-3d_{6}}{6\sqrt{2}}$\tabularnewline
\cline{2-5} 
 & $\alpha_{6^{*}}^{\phi}$ & \multicolumn{3}{c}{$d_{5}\rightarrow d_{8},d_{6}\rightarrow d_{9}$}\tabularnewline
\cline{2-5} 
 & $\alpha_{66^{*}}^{\phi}$ & \multicolumn{3}{c}{$d_{5}\rightarrow f_{3},d_{6}\rightarrow\tilde{f}_{3}$}\tabularnewline
 \cline{2-5} 

 & $\beta^{\pi}$ & $\sqrt{2}$ & $\frac{1}{2\sqrt{2}}$ & $\frac{1}{\sqrt{2}}$\tabularnewline

 & $\beta^{k}$ & $\frac{1}{\sqrt{2}}$ & $\frac{1}{\sqrt{2}}$ & $-\frac{1}{\sqrt{2}}$\tabularnewline

 & $\beta^{\eta}$ & $0$ & $\frac{3}{2\sqrt{2}}$ & $0$\tabularnewline
\midrule[1pt]
\multirow{9}{*}{$(m),(n),(o),(p)$} & $N_{3}^{\pi}$ & 6 & $\frac{3}{2}$ & $\frac{3}{2}$\tabularnewline
 & $N_{3}^{k}$ & 2 & $5$ & $5$\tabularnewline

 & $N_{3}^{\eta}$ & 0 & $\frac{3}{2}$ & $\frac{3}{2}$\tabularnewline
\cline{2-5} 

 & $N^{\pi}$ & $2$ & $\frac{3}{2}$ & $\frac{3}{2}$\tabularnewline

 & $N^{k}$ & $2$ & $1$ & $1$\tabularnewline

 & $N^{\eta}$ & $0$ & $\frac{3}{2}$ & $\frac{3}{2}$\tabularnewline
\cline{2-5} 

 & $O^{\pi}$ & $2$ & $\frac{3}{4}$ & $\frac{3}{4}$\tabularnewline

 & $O^{k}$ & $1$ & $\frac{5}{2}$ & $\frac{5}{2}$\tabularnewline
 & $O^{\eta}$ & $\frac{1}{3}$ & $\frac{1}{12}$ & $\frac{1}{12}$\tabularnewline
\bottomrule[1pt]\bottomrule[1pt]
\end{tabular}
\end{table}
\subsection{$B^{\mu}_{6^*}(p')\rightarrow B_{3}(p)+\gamma(q) $}
For the radiative transition $B^{\mu}_{6^*}\rightarrow B_{3}\gamma$, we give the explicit forms of the $G_1$ and $G_2$.
At the leading order, the $\mathcal L_{B\gamma}^{(2)}$ contributes to the form factor $G_1$,
\begin{equation}
G^{(2,\text{tree})}_{1}=-\frac{f_{4}}{2m_{N}}C^{(2)}_{\phi},
\end{equation}
where the superscript denotes that the value of $G_{1}$ comes from the $\mathcal O(p^2)$ tree diagram in Fig.~\ref{tree}. $G_2$ vanishes at this order.

At NLO, both the M1 and E2 transitions contribute to the radiative decay. The $\mathcal{L}^{(3)}_{B\gamma}$ contributes at the tree level and the from factors read,
\begin{eqnarray}
&&G_{1}^{(3,\text{tree})}=\frac{n_{2} \ell C_{\phi}^{(2)}}{4m_{N}^{2}}, G_{2}^{(3,\text{tree})}=\frac{n_{1}C_{\phi}^{(2)}}{2m_{N}^{2}}.
\end{eqnarray}
The loop diagrams $(a)$ and $(b)$ also contribute at this order,
\begin{eqnarray}
&&G_{1}^{(3,a-b)}=\left[-n_{5}^{\text{III}}(\delta_3,\delta_3-\ell)\frac{g_{2}g_{3}}{2F_{\phi}^{2}}+\left(n_{5}^{\text{III}}(0,-\ell)\frac{d-3}{d-1}+n_{1}^{\text{II}}(0,-\ell)\right)\frac{g_{4}g_{5}}{2F_{\phi}^{2}}\right]C_{\phi}^{(3)},\nonumber \\
&&G_{2}^{(3,a-b)}=\left[-\frac{g_{2}g_{3}}{2F_{\phi}^{2}}\left(2n_{2}^{\text{III}}(\delta_{3},\delta_{3}-\ell)+2n_{4}^{\text{II}}(\delta_{3},\delta_{3}-\ell)\right)+\frac{g_{4}g_{5}}{2F_{\phi}^{2}}\frac{d-3}{d-1}\left(2n_{2}^{\text{III}}(0,-\ell)+2n_{4}^{\text{II}}(0,-\ell)\right)\right]C_{\phi}^{(3)},\nonumber \\
\end{eqnarray}
where $n_{5}^{\text{III}}$, $n_{2}^{\text{III}}$ and $n_{4}^{\text{II}}$ are the finite parts of the loop integrals. 

At $\mathcal O(p^4)$, the analytical expressions of form factors coming from the tree diagram are 
\begin{eqnarray}
G_{1}^{(4,\text{tree})}(\Sigma_{c}^{+}\rightarrow\Lambda_{c}^{+})&=&0,\nonumber \\
G_{1}^{(4,\text{tree})}(\Xi_{c}^{'+}\rightarrow\Xi_{c}^{+})&=&\frac{2\left(h_{2}+h_{3}\right)-3h_{1}}{12M_N\sqrt{2}},\nonumber \\
G_{1}^{(4,\text{tree})}(\Xi_{c}^{'0}\rightarrow\Xi_{c}^{0})&=&-\frac{3h_{1}-2h_{2}+h_{3}}{12M_N\sqrt{2}}.
\end{eqnarray}
At $\mathcal O(p^4)$, the analytical expressions of form factors coming from the loop diagrams are 
\begin{eqnarray}
G_{1}^{(4,c)}&=&-\frac{f_4}{16m_{N}F_{\phi}^{2}}\delta^{\phi}\frac{m_{\phi}^{2}}{16\pi^{2}}\ln\frac{m_{\phi}^{2}}{\lambda^{2}},\\
G_{1}^{(4,d)}&=&-\frac{a_{3}}{4F_{\phi}^{2}m_{N}}\gamma^{\phi}\frac{m_{\phi}^{2}}{32\pi^{2}}\ln\frac{m_{\phi}^{2}}{\lambda^{2}}-\frac{a_{4}}{4F_{\phi}^{2}m_{N}}\gamma_{2}^{\phi}\frac{m_{\phi}^{2}}{32\pi^{2}}\ln\frac{m_{\phi}^{2}}{\lambda^{2}},\\
 G_{1}^{(4,e)} &=&-\alpha_{6^{*}}^{\phi}\frac{g_{4}g_{5}}{4F_{\phi}^{2}M_{N}}\frac{2\left(d^{2}-2d-3\right)}{(d-1)^{2}}\Lambda_{2}(-\ell,0),\\
G_{1}^{(4,f)}	&=&-\alpha_{6}^{\phi}\frac{g_{2}g_{3}}{4F_{\phi}^{2}M_{N}}\Lambda_{2}(\delta_{3}-\ell,\delta_{3})-\beta^{\phi}8\frac{g_{2}g_{4}f_{2}}{4F_{\phi}^{2}M_{N}}\Lambda_{2}(\delta_{3}-\ell,\delta_{2}),\\
G_{1}^{(4,g)}	&=&-\alpha_{6^{*}6}^{\phi}\frac{g_{2}g_{5}}{2F_{\phi}^{2}M_{N}}\left(\frac{3-d}{4}+\frac{2}{d-1}\right)\Lambda_{2}(\delta_{3}-\ell,0),\\
G_{1}^{(4,h)}	&=&-\alpha_{6^{*}6}^{\phi}\frac{g_{3}g_{4}}{8F_{\phi}^{2}M_{N}}\frac{d-5}{d-1}\Lambda_{2}(-\ell,\delta_{3})-\beta^{\phi}\frac{g_{4}^{2}}{8M_{N}F_{\phi}^{2}}\frac{d-5}{d-1}\Lambda_{2}(-\ell,\delta_{2}),\\
 G_{1}^{(4,m)}	&=&N_{3}^{\phi}\frac{g_{4}^{2}}{4F_{\phi}^{2}}(2-d)J_{2}'(-\ell)\times\frac{1}{2}G_{1}^{(2,\text{tree})},\\
G_{1}^{(4,n)}	&=&N_{3}^{\phi}\frac{g_{2}^{2}}{F_{\phi}^{2}}\frac{1-d}{4}J_{2}'(\delta_{3}-\ell)\times\frac{1}{2}G_{1}^{(2,\text{tree})},\\
G_{1}^{(4,o)}	&=&O^{\phi}\left(\frac{1-d}{4}+\frac{1}{d-1}\right)\frac{g_{5}^{2}}{F_{\phi}^{2}}J'_{2}(0)\times\frac{1}{2}G_{1}^{(2,\text{tree})},\\
G_{1}^{(4,p)}	&=&O^{\phi}\left[-\frac{g_{3}^{2}}{4F_{\phi}^{2}}J'_{2}(\delta_{3})-N^{\phi}\frac{g_{4}^{2}}{4F_{\phi}^{2}}J'_{2}(\delta_{2})\right]\times\frac{1}{2}G_{1}^{(2,\text{tree})}.
\end{eqnarray}

\subsection{$B^{\mu}_{6^*}(p')\rightarrow B_{6}(p)+\gamma(q) $}

The spin-$\frac{3}{2}$ sextet  decay into the spin-$\frac{1}{2}$ sextet through the M1 and E2 transitions. In this section, we show the analytical expressions of the form factors $G_1$ and $G_2$. 
Then, one can obtain the analytical expressions of the decay amplitudes and the transition magnetic moments using Eqs.~(\ref{tr13})-(\ref{tr16}).
 
At the leading order, the transition amplitude arises from the $\mathcal L^{(2)}_{B\gamma}$. The $G_2$ vanishes and the $G_1$ is
\begin{eqnarray}
G_{1}^{(2,\text{tree})}=-\frac{C_{6}^{(2)}f_{3}+\tilde{C}_{6}^{(2)}\tilde{f}_{3}}{2m_{N}},
\end{eqnarray}
where the $C_{6}^{(2)}$ and $\tilde{C}_{6}^{(2)}$ are the coefficients listed in Table~\ref{cg66}. 

At the next-to-leading order, both the tree and the loop diagrams contribute to the chiral corrections. The $\mathcal O(p^3)$ tree diagram arises from the $\mathcal L^{(3)}_{B\gamma}$ and the form factors are
\begin{eqnarray}
&&G_{1}^{(3,\text{tree})} =\left(m_{2}\frac{\ell}{4m_{N}}C_{6}^{(2)}+\widetilde{m}_{2}\frac{\ell}{4m_{N}}\tilde{C}_{6}^{(2)}\right), ~~~~~\ell=q\cdot v.\nonumber\\
&&G_{2}^{(3,\text{tree})}=\frac{m_{1}C_{6}^{(2)}+\widetilde{m}_{1}\tilde{C}_{6}^{(2)}}{2m_{N}^{2}}.
\end{eqnarray}

The analytical expressions of the $\mathcal O(p^3)$ loop diagrams (a), (b) in Fig.~\ref{allloop} are, 
\begin{eqnarray}
G_{1}^{(3,a)}	&=&\beta^{\phi}\frac{g_{5}g_{3}}{2F_{\phi}^{2}}\left(n_{5}^{\text{III}}(0,-\ell)\frac{d-3}{d-1}+n_{1}^{\text{II}}(0,-\ell)\right),\\
G_{1}^{(3,b)}	&=&-\beta^{\phi}\frac{g_{1}g_{3}}{2F_{\phi}^{2}}n_{5}^{\text{III}}(\delta_{3},\delta_{3}-\ell)-h^{\phi}\frac{g_{4}g_{2}}{2F_{\phi}^{2}}n_{5}^{\text{III}}(\delta_{2},\delta_{2}-\ell) \\
G_{2}^{(3,a)}&=&\beta^{\phi}\frac{g_{5}g_{3}}{F_{\phi}^{2}}\frac{d-3}{d-1}\left(n_{2}^{\text{III}}(0,-\ell)+n_{5}^{\text{II}}(0,-\ell)\right),\\
G_{2}^{(3,b)}&=&-\beta^{\phi}\frac{g_{1}g_{3}}{F_{\phi}^{2}}\left(n_{2}^{\text{III}}(\delta_{3},\delta_{3}-\ell)+n_{4}^{\text{II}}(\delta_{3},\delta_{3}-\ell)\right)-h^{\phi}\frac{g_{4}g_{2}}{F_{\phi}^{2}}\left(n_{2}^{\text{III}}(\delta_{2},\delta_{2}-\ell)+n_{4}^{\text{II}}(\delta_{2},\delta_{2}-\ell)\right) \nonumber\\
\end{eqnarray}
where $\beta^{\phi}$, $h^{\phi}$ and the following $\delta_{\phi}$, $\theta_\phi$ and so on are the coefficients as listed in Table~\ref{cg66}. 

At NNLO, the form factors from the $\mathcal O(p^4) $ loop diagrams and the tree diagram are,
\begin{eqnarray}
G_{1}^{(4,c)}&=&-\delta_{66}^{\phi}\frac{1}{16M_{N}F_{\phi}^{2}}\frac{m_{\phi}^{2}}{16\pi^{2}}\ln\frac{m_{\phi}^{2}}{\lambda^{2}},\\
G_{1}^{(4,d)}&=&-\gamma_{6}^{\phi}\frac{a_{5}}{4M_{N}F_{\phi}^{2}}\frac{m_{\phi}^{2}}{32\pi^{2}}\ln\frac{m_{\phi}^{2}}{\lambda^{2}},\\
G_{1}^{(3,e)}	&=&-\theta_{2}^{\phi}\frac{g_{3}g_{5}}{4M_{N}F_{\phi}^{2}}\frac{2\left(d^{2}-2d-3\right)}{(d-1)^{2}}\Lambda_{2}(-\ell,0),\\
G_{1}^{(3,f)}	&=&-\theta_{1}^{\phi}\frac{g_{1}g_{3}}{4M_{N}F_{\phi}^{2}}\Lambda_{2}(\delta_{3}-\ell,\delta_{3})-\theta_{3}^{\phi}\frac{g_{2}g_{4}}{4M_{N}F_{\phi}^{2}}\Lambda_{2}(\delta_{3}-\ell,\delta_{3})\nonumber \\
	&~~&-\theta_{4}^{\phi}8\frac{g_{1}g_{4}f_{2}}{4M_{N}F_{\phi}^{2}}\Lambda_{2}(\delta_{3}-\ell,\delta_{2})-\theta_{4}^{\phi}8\frac{g_{1}g_{4}f_{2}}{4M_{N}F_{\phi}^{2}}\Lambda_{2}(\delta_{2}-\ell,\delta_{3}),\\
G_{1}^{(3,g)}	&=&-\theta^{\phi}\frac{g_{1}g_{5}}{2M_{N}F_{\phi}^{2}}\left(\frac{3-d}{4}+\frac{2}{d-1}\right)\Lambda_{2}(\delta_{3}-\ell,0)\nonumber\\
&~~&-\theta_{4}^{\phi}\frac{g_{2}g_{5}f_{4}}{2M_{N}F_{\phi}^{2}}\left(\frac{3-d}{4}+\frac{2}{d-1}\right)\Lambda_{2}(\delta_{2}-\ell,0),\\
G_{1}^{(3,h)}	&=&-\theta^{\phi}\frac{g_{3}^{2}}{8M_{N}F_{\phi}^{2}}\left(\frac{d-5}{d-1}\right)\Lambda_{2}(\delta_{3},-\ell)-\theta_{4}^{\phi}\frac{g_{3}g_{4}f_{4}}{8M_{N}F_{\phi}^{2}}\left(\frac{d-5}{d-1}\right)\Lambda_{2}(\delta_{2},-\ell),\\
 G_{1}^{(4,m)}	&=&O^{\phi}\frac{g_{3}^{2}}{4F_{\phi}^{2}}J_{2}'(-\ell)(2-d)\times\frac{1}{2}G_{1}^{(2)},\\
G_{1}^{(4,n)}	&=&\left[O^{\phi}\frac{g_{1}^{2}}{F_{\phi}^{2}}\frac{1-d}{4}J_{2}'(\delta_{3}-\ell)+N^{\phi}\frac{g_{2}^{2}}{F_{\phi}^{2}}\frac{1-d}{4}J_{2}'(\delta_{2}-\ell)\right]\times\frac{1}{2}G_{1}^{(2)},\\
G_{1}^{(4,o)}	&=&O^{\phi}\frac{g_{5}^{2}}{F_{\phi}^{2}}\left(\frac{1-d}{4}+\frac{1}{d-1}\right)J_{2}'(0)\times\frac{1}{2}G_{1}^{(2)},\\
G_{1}^{(4,p)}	&=&-O^{\phi}\frac{g_{3}^{2}}{4F_{\phi}^{2}}J_{2}'(\delta_{3})-N^{\phi}\frac{g_{4}^{2}}{4F_{\phi}^{2}}J_{2}'(\delta_{2})\times\frac{1}{2}G_{1}^{(2)},\\
G_1{1}^{(4,\text{tree})}&=&-{D_4\over 4M_N}.
\end{eqnarray}

\begin{table}[!htp]
 \caption{The coefficients for the tree and loop diagrams that contribute to the amplitudes in the radiative decays $B^{\mu}_{6^*}\rightarrow B_6\gamma$ up to NNLO. }\label{cg66}
\begin{tabular}{c|c|cccccc}
\toprule[1pt]\toprule[1pt]
 & & $\Sigma_{c}^{*++}\rightarrow\Sigma_{c}^{++}\gamma$ & $\Sigma_{c}^{*+}\rightarrow\Sigma_{c}^{+}\gamma$ & $\Sigma_{c}^{*0}\rightarrow\Sigma_{c}^{0}\gamma$ & $\Xi_{c}^{*+'}\rightarrow\Xi_{c}^{+'}\gamma$ & $\Xi_{c}^{*0}\rightarrow\Xi_{c}^{0'}\gamma$ & $\Omega_{c}^{*0}\rightarrow\Omega_{c}^{0}\gamma$\tabularnewline
\midrule[1pt]
\multirow{2}{*}{$\mathcal{O}(p^{2}),\mathcal{O}(p^{3})$ Tree } & $C_{6}^{(2)}$ & $\frac{2}{3}$ & $\frac{1}{6}$ & $-\frac{1}{3}$ & $\frac{1}{6}$ & $-\frac{1}{3}$ & $-\frac{1}{3}$\tabularnewline
 & $\tilde{C}_{6}^{(2)}$ & $1$ & $1$ & $1$ & $1$ & $1$ & $1$\tabularnewline
\midrule[1pt]
\multirow{4}{*}{$(a),(b)$} & $\beta^{\pi}$ & $1$ & $0$ & $-1$ & $\frac{1}{2}$ & $-\frac{1}{2}$ & $0$\tabularnewline

 & $\beta^{k}$ & $1$ & $\frac{1}{2}$ & $0$ & $0$ & $-\frac{1}{2}$ & $-1$\tabularnewline
\cline{2-8} 
 & $h^{\pi}$ & $2$ & $0$ & $-2$ & $1$ & $-1$ & $0$\tabularnewline

 & $h^{k}$ & $2$ & $1$ & $0$ & $0$ & $-1$ & $-2$\tabularnewline
\midrule[1pt]
\multirow{4}{*}{$(c),(d)$} & $\delta_{66}^{\pi}$ & $-4f_{3}$ & $0$ & $4f_{3}$ & $-2f_{3}$ & $2f_{3}$ & $0$\tabularnewline

 & $\delta_{66}^{k}$ & $-4f_{3}$ & $-2f_{3}$ & $0$ & $0$ & $2f_{3}$ & $4f_{3}$\tabularnewline
\cline{2-8} 
 & $\gamma_{6}^{\pi}$ & $2$ & $0$ & $-2$ & $1$ & $-1$ & $0$\tabularnewline

 & $\gamma_{6}^{k}$ & $2$ & $1$ & $0$ & $0$ & $-1$ & $-2$\tabularnewline
\midrule[1pt]
\multirow{9}{*}{$(e),(f),(g),(h)$} & $\theta^{\pi}$ & $2\tilde{f_{3}}+\frac{5f_{3}}{6}$ & $\frac{1}{3}\left(6\tilde{f_{3}}+f_{3}\right)$ & $2\tilde{f_{3}}-\frac{f_{3}}{6}$ & $\frac{1}{8}\left(6\tilde{f_{3}}-f_{3}\right)$ & $\frac{3\tilde{f_{3}}}{4}$ & $0$\tabularnewline

 & $\theta^{k}$ & $\tilde{f_{3}}+\frac{f_{3}}{6}$ & $\tilde{f_{3}}-\frac{f_{3}}{12}$ & $\tilde{f_{3}}-\frac{f_{3}}{3}$ & $\frac{5}{12}\left(6\tilde{f_{3}}+f_{3}\right)$ & $\frac{5\tilde{f_{3}}}{2}-\frac{7f_{3}}{12}$ & $2\tilde{f_{3}}-\frac{f_{3}}{6}$\tabularnewline

 & $\theta^{\eta}$ & $\frac{1}{9}\left(3\tilde{f_{3}}+2f_{3}\right)$ & $\frac{1}{18}\left(6\tilde{f_{3}}+f_{3}\right)$ & $\frac{1}{9}\left(3\tilde{f_{3}}-f_{3}\right)$ & $\frac{1}{72}\left(6\tilde{f_{3}}+f_{3}\right)$ & $\frac{1}{36}\left(3\tilde{f_{3}}-f_{3}\right)$ & $\frac{4}{9}\left(3\tilde{f_{3}}-f_{3}\right)$\tabularnewline
 \cline{2-8} 
 & $\theta_{3}^{\pi}$ & $\frac{1}{3}\left(d_{2}+6d_{3}\right)$ & $\frac{1}{3}\left(d_{2}+6d_{3}\right)$ & $\frac{1}{3}\left(d_{2}+6d_{3}\right)$ & $\frac{1}{4}\left(6d_{3}-d_{2}\right)$ & $\frac{3d_{3}}{2}$ & $0$\tabularnewline

 & $\theta_{3}^{k}$ & $\frac{1}{3}\left(d_{2}+6d_{3}\right)$ & $2d_{3}-\frac{d_{2}}{6}$ & $2d_{3}-\frac{2d_{2}}{3}$ & $\frac{d_{2}}{6}+d_{3}$ & $\frac{d_{2}}{6}+d_{3}$ & $4d_{3}-\frac{d_{2}}{3}$\tabularnewline

 & $\theta_{3}^{\eta}$ & $0$ & $0$ & $0$ & $\frac{1}{4}\left(d_{2}+6d_{3}\right)$ & $\frac{1}{2}\left(3d_{3}-d_{2}\right)$ & $0$\tabularnewline

\cline{2-8} 
 & $\theta_{4}^{\pi}$ & $-1$ & $0$ & $1$ & $\frac{1}{4}$ & $\frac{1}{2}$ & $0$\tabularnewline

 & $\theta_{4}^{k}$ & $-1$ & $-\frac{1}{2}$ & $0$ & $-\frac{1}{2}$ & $\frac{1}{2}$ & $1$\tabularnewline

 & $\theta_{4}^{\eta}$ & $0$ & $0$ & $0$ & $-\frac{1}{4}$ & $0$ & $0$\tabularnewline
 \cline{2-8}
 & $\theta_1^{\phi}$ & \multicolumn{6}{c}{$f_{3}\rightarrow d_{5},\tilde f_3 \rightarrow d_{6}$}\tabularnewline
\cline{2-8}
 & $\theta_2^{\phi}$ & \multicolumn{6}{c}{$f_{3}\rightarrow d_{9},\tilde f_3 \rightarrow d_{9}$}\tabularnewline

\midrule[1pt]
\multirow{6}{*}{$(m),(n),(o),(p)$} & $N^{\pi}$ & $2$ & $2$ & $2$ & $\frac{3}{2}$ & $\frac{3}{2}$ & $0$\tabularnewline
 & $N^{k}$ & $2$ & $2$ & $2$ & $1$ & $1$ & $4$\tabularnewline

 & $N^{\eta}$ & $0$ & $0$ & $0$ & $\frac{3}{2}$ & $\frac{3}{2}$ & $0$\tabularnewline
\cline{2-8} 
 & $O^{\pi}$ & $2$ & $2$ & $2$ & $\frac{3}{4}$ & $\frac{3}{4}$ & $0$\tabularnewline

 & $O^{k}$ & $1$ & $1$ & $1$ & $\frac{5}{2}$ & $\frac{5}{2}$ & $2$\tabularnewline

 & $O^{\eta}$ & $\frac{1}{3}$ & $\frac{1}{3}$ & $\frac{1}{3}$ & $\frac{1}{12}$ & $\frac{1}{12}$ & $\frac{4}{3}$\tabularnewline
\midrule[1pt]
$\mathcal{O}(p^{4})$ Tree & $D_{4}$ & $0$ & $0$ & $0$ & $\frac{l_{1}}{2}+\frac{l_{2}}{3}$ & $\frac{l_{1}}{2}-\frac{l_{2}}{6}$ & $l_{1}-\frac{l_{2}}{3}$\tabularnewline
\bottomrule[1pt]\bottomrule[1pt]
\end{tabular}
\end{table}

\subsection{The U-spin symmetry in the analytical expressions}\label{relation}

For the transitions $B_{6}\rightarrow B_{\bar{3}}\gamma$ and $B^{\mu}_{6^*}\rightarrow B_{\bar{3}}\gamma$, the form factors and the transition magnetic moments of the heavy baryons completely come from the dynamics of the two inner light quarks. The contributions from the two light quarks are destructive, which is clearer in the quark model as listed in Appendix~\ref{QM}. 

In the neutral decays $\Xi_c^{'0} \rightarrow \Xi^0_c\gamma$ and $\Xi_c^{*0} \rightarrow \Xi^0_c \gamma$, the two light quarks are $s$ and $d$. In the $\mathcal O(p^2)$ and $\mathcal O(p^3)$ tree diagrams, their contributions cancel out because their masses and charges are the same in the SU(3) flavor symmetry. The coefficient $C^{(2)}$ for the two tree diagrams vanishes. Then, the decay amplitudes totally come from the chiral corrections of the loop diagrams (a) and (b) in Fig.~\ref{allloop} up to NLO. In these diagrams, the coefficients of the $\pi$ and $K$ loops are opposite as illustrated in Table~\ref{cg63}. In the exact SU(3) flavor symmetry, the masses and decay constants of the $\pi$, $K$ and $\eta$ are the same. Then the $\pi$ loop and $K$ loop in (a) or (b) cancel out exactly. In this work, we introduce the breaking effects of the U-spin symmetry through the masses and decay constants of the $\pi$ and $K$ mesons in the loops. The decay widths are nonvanishing.

At NNLO, the above conclusion also holds. The wave function renormalization diagrams do not contribute since the amplitudes of the $\mathcal O(p^2)$ tree diagrams vanish. The $\pi$ loop and $K$ loop in the (c) and (d) diagrams cancel out with each other. In the (e)-(h) diagrams,  the sum of the $\pi$ loop, $K$ loop and $\eta$ loop cancel out. The $\mathcal L^{(4)}_{B\gamma}$ can be absorbed into the $\mathcal L^{(2)}_{B\gamma}$ in the exact SU(3) flavor symmetry. 

In conclusion, the decay widths of the $\Xi_c^{'0} \rightarrow \Xi^0_c\gamma$ and $\Xi_c^{*0} \rightarrow \Xi^0_c \gamma$ totally arise from the U-spin symmetry breaking effects up to NNLO.

Another manifestation of the U-spin symmetry is the relations among the coefficients of the charged radiative decays. If we exchange the $s$ quark and $d$ quark, the heavy baryons transform as $\Sigma^+_c\rightarrow \Xi^{'+}_c$, $\Lambda^+_c \rightarrow \Xi^+_c$. One obtains
\begin{eqnarray}\label{re}
 X^{\pi(K)}(\Sigma^{(*)+}_c\rightarrow \Lambda^+_c\gamma)=X^{K(\pi)}(\Xi^{{'(*)}+}_c\rightarrow \Xi^+_c\gamma),
\end{eqnarray}
where $X$ denotes the coefficients $C^{(3)}$, $\delta^\phi$ and so on for the diagrams (a)-(d).

For the radiative decays $B^{\mu}_{6^*}\rightarrow B_6 \gamma$, there are similar relations between the coefficients as Eq.~(\ref{re}), 
\begin{eqnarray}
 X^{\pi(K)}(\Sigma^{*+}_c\rightarrow \Sigma^+_c\gamma)=X^{K(\pi)}(\Xi^{*+}_c\rightarrow \Xi^{'+}_c\gamma), ~~~X^{\pi(K)}(\Sigma^{*0}_c\rightarrow \Sigma^0_c\gamma)=X^{K(\pi)}(\Omega^{*0}_c\rightarrow \Omega^{0}_c\gamma),
\end{eqnarray}
where $X$ denotes the coefficients in Table~\ref{cg66} for the diagrams (a)-(d).

We also find some relations between the form factors in Table~\ref{cg66}.
Up to NLO, one obtains similar relations as those in Ref.~\cite{Banuls:1999br},
\begin{eqnarray}
G_{M1}(\Sigma^{*++}_c\rightarrow \Sigma^{++}_c \gamma)+G_{M1}(\Sigma^{*0}_c\rightarrow \Sigma^{0}_c \gamma)&=&2G_{M1}(\Sigma^{*+}_c\rightarrow\Sigma^{+}_c \gamma).\label{re6p4}\\
G_{M1}(\Sigma^{*++}_c\rightarrow \Sigma^{++}_c \gamma)+2G_{M1}(\Xi^{*0}_c\rightarrow \Xi^{'0}_c\gamma)&=&G_{M1}(\Sigma^{*0}_c\rightarrow \Sigma^{0}_c \gamma)+2G_{M1}(\Xi^{*+}_c\rightarrow \Xi^{'+}_c\gamma)\nonumber\\
 &=&G_{M1}(\Omega^{*0}_c\rightarrow \Omega^{0}_c \gamma)+2G_{M1}(\Sigma^{*+}_c\rightarrow \Sigma^+_c).\label{re6p3}
\end{eqnarray}
The $G_{E2}$ also satisfies the same relationships. Up to NNLO, Eq.~(\ref{re6p4}) still holds. Eq.~(\ref{re6p3}) is destroyed by the $\mathcal O(p^4)$ loop diagrams. In the calculation of the transition magnetic moments and the amplitudes, we use the baryon masses as listed in Table~\ref{mass}. 

\begin{table}
\caption{The masses of the heavy baryons in the unit of MeV. The masses without special notations are from Ref.~\cite{Tanabashi:2018oca}. The ${\dagger}$ represents that the mass of the corresponding state is still absent. Then we estimate it with the average mass of the other states in the same isospin multiplet. }\label{mass}
\begin{tabular}{ccc|cccccc}
\toprule[1pt] \toprule[1pt] 
\multirow{2}{*}{$\Lambda_{c}^{+}$} & \multirow{2}{*}{$\text{\ensuremath{\Xi}}_{c}^{+}$} & \multirow{2}{*}{$\text{\ensuremath{\Xi}}_{c}^{0}$} & $\Sigma_{c}^{++}$ & $\Sigma_{c}^{+}$ & $\Sigma_{c}^{0}$ & $\Xi_{c}^{'+}$ & $\Xi_{c}^{'0}$ & $\Omega_{c}^{0}$\tabularnewline
%\cline{4-9} 
 & & & $2453.97$ & $2452.9$ & $2453.75$ & $2577.4$ & $2578.8$ & $2695.2$\tabularnewline
\midrule[1pt]
\multirow{2}{*}{$2286.46$} & \multirow{2}{*}{$2467.87$} & \multirow{2}{*}{$2470.87$} & $\Sigma_{c}^{*++}$ & $\Sigma_{c}^{*+}$ & $\Sigma_{c}^{*0}$ & $\Xi_{c}^{*+}$ & $\Xi_{c}^{*0}$ & $\Omega_{c}^{*0}$\tabularnewline
%\cline{4-9} 
 & & & $2518.41$ & $2517.5$ & $2518.48$ & $2645.53$ & $2646.32$ & $2765.9$\tabularnewline
\midrule[1pt]
\multirow{2}{*}{$\Lambda_{b}^{0}$} & \multirow{2}{*}{$\Xi_{b}^{0}$} & \multirow{2}{*}{$\Xi_{b}^{-}$} & $\Sigma_{b}^{+}$ & $\Sigma_{b}^{0}$ & $\Sigma_{b}^{-}$ & $\Xi_{b}^{'0}$ & $\Xi_{b}^{'-}$ & $\Omega_{b}^{-}$\tabularnewline
%\cline{4-9} 
 & & & $5811.3$ & $5813.4^{\dagger}$ & $5815.5$ & $5935.02^{\dagger}$ & $5935.02$ & $6046.1$\tabularnewline
\midrule[1pt]
\multirow{2}{*}{$5619.6$} & \multirow{2}{*}{$5791.9$} & \multirow{2}{*}{$5794.5$} & $\Sigma_{b}^{*+}$ & $\Sigma_{b}^{*0}$ & $\Sigma_{b}^{*-}$ & $\Xi_{b}^{*0}$ & $\Xi_{b}^{*-}$ & $\Omega_{b}^{*-}$\cite{Jenkins:1996rr}\tabularnewline
%\cline{4-9} 
 & & & $5832.1$ & $5833.6^{\dagger}$ & $5835.1$ & $5949.8$ & $5955.33$ & $6083.2$\tabularnewline
\bottomrule[1pt] \bottomrule[1pt] 
\end{tabular}
 \end{table}

\section{The independent LECs in the heavy quark limit}\label{sec4}
In previous works, we have calculated the magnetic moments of the spin-$1\over 2$ and spin-$3\over 2$ heavy baryons up to NNLO~\cite{Wang:2018gpl,Meng:2018gan}. There are many common LECs  for the magnetic moments and the radiative decay amplitudes. Thus, we perform the numerical analysis for the radiative decay widths together with the magnetic moments up to NNLO. 

At the leading order, there are ten LECs: the $d_{2,3,5,6,8,9}$, $f_{2,3,4}$ and $\tilde f_3$. At NLO, the magnetic moments and the decay amplitudes contain nine LECs, including five axial coupling constants $g_{1-5}$ in the $\mathcal O(p^3)$ loop diagrams and three LECs in the $\mathcal O (p^3)$ tree diagrams: $n_{1}$, $m_{1}$ and $\tilde{m}_{1}$.

 At NNLO, there are eight LECs $a_{1-5}$, $d_1$, $d_4$ and $d_7$ in the loop diagrams. In the tree diagrams, there are eight LECs $c_{1-3}$, $h_{1-3}$ and $l_{1,2}$ for the radiative transition and five LECs $s_{2-6}$ for the magnetic moments. In general, these LECs should have been estimated with the experiment data as input. So far, there are no experiment data. As a compromise, we use the data from the lattice QCD simulation as input, which is listed in Table~\ref{lattice}. One notices that the number of the lattice QCD data is still smaller than that of the LECs. In the following section, we use the heavy quark symmetry to reduce the number of the LECs.

\begin{table}
 \caption{The data from the lattice QCD simulation~\cite{Bahtiyar:2015sga,Bahtiyar:2016dom,Can:2013tna,Can:2015exa}. The value of $G_{M1}(\Omega_{c}^{*0}\rightarrow\Omega_{c}^{0}\gamma)$ is derived from  Ref.~\cite{Bahtiyar:2016dom}. The magnetic moment is in the unit of the nuclear magneton.  The superscript $\ddag$ denotes that the corresponding data is treated as input.}\label{lattice}
\begin{tabular}{c|c|c|c|c|cccc}
\toprule[1pt] \toprule[1pt] 
$\mu^{\ddagger}_{\Xi_{c}^{+}}$ & $\ensuremath{\mu_{\Xi_{c}^{0}}}$ & $\ensuremath{\mu_{\Xi_{c}^{'+}}}$ & $\ensuremath{\mu_{\Xi_{c}^{'0}}}$ & $\ensuremath{\mu^{\ddag}_{{\Xi_{c}^{'+}}\rightarrow{{\Xi_{c}^{+}}}\gamma}}$ & $\ensuremath{\mu_{\Xi_{c}^{'0}\rightarrow \Xi_{c}^{0}\gamma} }$\tabularnewline
\midrule[1pt] 
$0.235(25)$ & $\ensuremath{0.192(17)}$ & $\ensuremath{0.315(141)}$ & $\ensuremath{-0.599(71)}$ & $0.729(103)$ & $\ensuremath{0.009(13)}$\tabularnewline
\midrule[1pt] 
$\ensuremath{\mu^{\ddagger}_{\Sigma_{c}^{++}}}$ & $\ensuremath{\mu_{\Sigma_{c}^{0}}}$ & $\ensuremath{\mu^{\ddagger}_{\Omega_{c}^{0}}}$ & ${\mu^{\ddagger}_{\Omega_{c}^{*0}}}$ &\multicolumn{2}{c}{$G_{M1}(\Omega_{c}^{*0}\rightarrow\Omega_{c}^{0}\gamma)$} \tabularnewline
\midrule[1pt] 
$1.499(202)$ & $-0.875(103)$ & $-0.688(31)$ & $-0.730(23)$ & $G^q_{M1}(0)=0.671$ &$G^c_{M1}(0)=0.145$ \tabularnewline%$G_{E2}(0)=−0.012(62)$
\bottomrule[1pt]\bottomrule[1pt] 
\end{tabular}
\end{table}

\subsection{The heavy quark symmetry}
Besides the Lagrangians in Section~\ref{sec2}, the magnetic moments up to NNLO involve the following Lagrangians,
\begin{eqnarray}
{\mathcal{L}}_{MB}^{(2)}&=&\frac{d_{1}}{2M_{N}}\text{Tr}(\bar{B}_{\bar{3}}[S^{\mu},S^{\nu}][u_{\mu},u_{\nu}]B_{\bar{3}})+\frac{d_{4}}{M_{N}}\text{Tr}(\bar{B}_{6}[S^{\mu},S^{\nu}][u_{\mu},u_{\nu}]B_{6})\nonumber \\
&~~&+\frac{d_{7}}{M_{N}}\text{Tr}(\bar{B}_{6^{*}}^{\mu}[u_{\mu},u_{\nu}]B_{6^{*}}^{\nu}).\label{bpp2}\\
\mathcal{L}_{B\gamma}^{(4)} &=&-\frac{is_{2}}{4M_N}\text{Tr}(\bar{B}_{6}[S^{\mu},S^{\nu}]\chi_{+}B_{6})\text{Tr}(f_{\mu\nu}^{+})-\frac{is_{3}}{4m_{N}}\text{Tr}(\bar{B}_{6}^{ab}[S^{\mu},S^{\nu}]\{\chi_{+},\tilde{f}_{\mu\nu}^{+}\}_{ab}^{ij}B_{6ij}^{\nu})\nonumber\\
&~~&+\frac{is_{7}}{4m_{N}}\text{Tr}(\bar{B}_{6^*}^{\mu}\chi_{+}B_{6^{*}}^{\nu})\text{Tr}(\tilde{f}_{\mu\nu}^{+})+\frac{is_{8}}{4m_{N}}\text{Tr}(\bar{B}_{6^*}^{\mu}\{\chi_{+},\tilde{f}{}_{\mu\nu}^{+}\}_{ab}^{ij}B_{6^*}^{\nu}).\label{mp4}
%&-&\frac{1}{2}\frac{is_{6}}{4M_N}\text{Tr}(\bar B_{\bar{3}}[S^{\mu},S^{\nu}]\chi_{+}B_{\bar{3}})\text{Tr}(\tilde{f}_{\mu\nu}^{+})\nonumber\\
\end{eqnarray}

In the heavy quark limit,the spin-$1\over 2$ and spin-$3\over 2$ sextets are in the same multiplet. They can be described by a superfield~\cite{Falk:1991nq},
\begin{eqnarray}
&&\psi^{\mu}=B_{6^{*}}^{\mu}-\sqrt{\frac{1}{3}}(\gamma^{\mu}+v^{\mu})\gamma_{5}B_6,\\
&&\bar{\text{\ensuremath{\psi}}}{}_{\mu}=\bar{B}_{6^{*}}^{\mu}+\sqrt{\frac{1}{3}}\bar{B}_6\gamma_{5}(\gamma_{\mu}+v_{\mu}).
\end{eqnarray}
With the superfield, we construct the Lagrangians, the $\kappa_{1-8}$ terms, to reduce the number of the LECs.

The $\mathcal O (p^2)$ Lagrangians that contribute to the radiative decays read

\begin{eqnarray}
&\mathcal{L}_{HQSS}^{(2)}=i\frac{\kappa_{1}}{M_{N}}\text{Tr}(\bar{\psi}^{\mu}\tilde{f}_{\mu\nu}^{+}\psi^{\nu})+\frac{\kappa_{2}}{M_{N}}\epsilon_{\mu\nu\alpha\beta}\text{Tr}(\bar{\psi}^{\mu}\tilde{f}^{\alpha\beta}v^{\nu}B_{\bar 3}),\\
&\mathcal{L}_{QB}^{(2)}=\frac{\kappa_{3}}{M_{N}}\text{Tr}(\bar{\psi}^{\lambda}\sigma_{\mu\nu}\psi_{\lambda})\text{Tr}(f^{\mu\nu+}),
\end{eqnarray}
where the subscript ``$QB$'' represents the breaking effect of the heavy quark spin symmetry. Combining the two equations with Eq.~(\ref{trp2}), we reduce the seven LECs, $d_{5,6,8,9}$, $f_{2,3}$ and $\tilde{f}_{3}$, to three independent LECs, $\kappa_{1,2,3}$,
\begin{eqnarray}
&d_{5}=-\frac{8}{3}\kappa_{1},\,\,\,d_{8}=-2\kappa_{1},\,\,\,f_{3}=4\sqrt{\frac{1}{3}}\kappa_{1},\\
& f_{4}=8\kappa_{2},\,\,\,f_{2}=\frac{1}{\sqrt{3}}\kappa_{2},\\
& d_{6}=\frac{8}{3}\kappa_{3},\,\,\,d_{9}=-4\kappa_{3},\,\,\,\tilde{f}_{3}=-\frac{16}{\sqrt{3}}\kappa_{3}.
\end{eqnarray}
The Lagrangian that introduces the vertex $B\phi\phi$ at $\mathcal O(p^2)$ is
\begin{eqnarray}
\mathcal{L}_{B\phi\phi}^{(2)}=\frac{\kappa_{4}}{M_{N}}\text{Tr}(\bar{\psi}^{\mu}[u_{\mu},u_{\nu}]\psi^{\nu})+\frac{i\kappa_{5}}{M_{N}}\epsilon^{\sigma\mu\nu\rho}\text{Tr}(\bar{B}_{\bar{3}}[u_{\mu},u_{\nu}]v_{\rho}\psi_{\sigma})+\frac{i\kappa_{6}}{M_{N}}\text{Tr}(\bar{B}_{\bar{3}ab}\epsilon^{\sigma\mu\nu\rho}u_{i\mu}^{b}u_{j\nu}^{a}v_{\rho}\psi_{\sigma}^{ij}).
\end{eqnarray}
The LECs $d_{1,4,7}$ in Eq.~(\ref{bpp2}) and $a_{1,2,3,4,5}$ in Eq.~(\ref{diph}) are reduced to three independent LECs $\kappa_{4,5,6}$ as follows, 
\begin{eqnarray}
&a_{5}=-2\sqrt{\frac{1}{3}}\kappa_{4},\,\,\,d_{4}=\frac{2}{3}\kappa_{4},\,\,\,d_{7}=\kappa_{4},\\
&a_{3}=2\kappa_{5},\,\,\,a_{1}=4\sqrt{\frac{1}{3}}\kappa_{5},\\
&a_{4}=4\kappa_{6},\,\,\,a_{2}=2\sqrt{\frac{1}{3}}\kappa_{6}. 
\end{eqnarray}
 At $\mathcal O(p^4)$, the Lagrangian reads
\begin{eqnarray}
\mathcal{L}_{B\gamma}^{(4)}=\frac{i\kappa_{7}}{m_{N}}\text{Tr}(\bar{\psi}_{\mu}^{ab}\{\chi_{+},\tilde{f}_{\mu\nu}^{+}\}_{ab}^{ij}\psi_{ij}^{\nu})+\frac{\kappa_{8}}{m_{N}}\text{Tr}(\bar{\psi}^{\lambda}\chi_{+}\sigma^{\mu\nu}\psi_{\lambda})\text{Tr}(f_{\mu\nu}^{+}),
\end{eqnarray}
The LECs $s_{12,3,7,8}$ in Eq.~(\ref{mp4}) and $l_{1,2}$ in Eq.~(\ref{trp4}) are related to two independent LECs $\kappa_{7,8}$,
\begin{eqnarray}
&l_{2}=8\sqrt{\frac{1}{3}}\kappa_{7},\,\,\,s_{3}=-\frac{8}{3}\kappa_{7},~~~s_8=4\kappa_{7},\\
&l_{1}=-32\sqrt{\frac{1}{3}}\kappa_{8},~~~s_{2}=\frac{8}{3}\kappa_{8},~~~s_7=8\kappa_{8}.
\end{eqnarray}

In conclusion, up to NNLO, the LECs for the magnetic moments and the radiative decay amplitudes of the heavy baryons can be expressed by eleven independent LECs: $\kappa_{1-8}$, $n_{1}$, $m_{1}$ and $\tilde m_{1}$ in the heavy quark limit.

\section{NUMERICAL RESULTS AND DISCUSSIONs }\label{sec5}
\subsection{The radiative decays from the sextet to the antitriplet charmed baryons}

For the radiative transitions $B_6\rightarrow B_{\bar 3}\gamma$ and  $B^\mu_{6^*}\rightarrow B_{\bar 3}\gamma$, we calculate the numerical results up to NLO. The numerical results are listed in Table~\ref{63reults}.  Their analytical expressions contain three unknown coefficients $f_2$, $f_4$, and $n_1$. The $f_2$ is estimated using $\mu({\Xi_{c}^{'+}\rightarrow{\Xi_{c}^{+}}\gamma})$ from lattice QCD simulation and $f_4$ is related to  $f_2$ through $\kappa_2$ in the heavy quark limit. The $n_1$ contributes to the $G_2$ form factor, which are important for the $G_{E2}$ and has little influence on $G_{M1}$. The radiative decay width mainly  arises from the M1 transition. Then we calculate the decay width without the $G_2$ contribution.

\begin{table}
 \caption{The transition magnetic moment  and the decay width for the radiative transition $B_6/B^{\mu}_{6^*}\rightarrow B_{\bar 3}\gamma$ in the charmed baryon sector. $\mu$ is in the unit of nuclear magneton. The superscript $\ddag$ denotes that the corresponding data is used as input.}\label{63reults}
 \begin{tabular}{l|ccc|c}
\toprule[1pt] \toprule[1pt] 
\multirow{2}{*}{Channel} & \multicolumn{3}{c|}{$\mu~(\mu_N)$}  & \multirow{2}{*}{$\Gamma$~(keV)}\tabularnewline
\cline{2-4} 
 & $\mathcal{O}(p)$ & $\mathcal{O}(p^{2})$ & Total & \tabularnewline
\midrule[1pt]
${\Sigma_{c}^{+}\rightarrow\Lambda_{c}^{+}\gamma}$ & $-2.70$ & $1.32$ & $-1.38$ & $65.6$ \tabularnewline
 $ {\Xi_{c}^{'+}\rightarrow\text{\ensuremath{\Xi}}_{c}^{+}\gamma}$ & $-2.70$ & $1.97$ & $0.73^{\ddag}$ & $5.43$\tabularnewline
 $ {\Xi_{c}^{'0}\rightarrow\text{\ensuremath{\Xi}}_{c}^{0}\gamma}$ & $0$ & $0.22$ & $0.22$ & $0.46$ \tabularnewline
\midrule[1pt]
$ {\Sigma_{c}^{*+}\rightarrow\Lambda_{c}^{+}\gamma}$ & $3.91$ & $-1.91$ & $2.00$ & $161.6$ \tabularnewline
$ {\Xi_{c}^{*+}\rightarrow\text{\ensuremath{\Xi}}_{c}^{+}\gamma}$ & $3.88$ & $-2.83$ & $1.05$ & $21.6$ \tabularnewline
 $ {\Xi_{c}^{*0}\rightarrow\text{\ensuremath{\Xi}}_{c}^{'0}\gamma}$ & $0$ & $-0.31$ & $-0.31$ & $1.84$ \tabularnewline
\bottomrule[1pt] \bottomrule[1pt] 
\end{tabular}
\end{table}

The radiative decay amplitudes of $\Xi_c^{'0} \rightarrow \Xi^0_c\gamma$ and $\Xi_c^{*0} \rightarrow \Xi^0_c \gamma$ completely come from the loops (a) and (b) up to NLO as illustrated in Section~\ref{relation}. The amplitudes of the two loops only involve $g_{1-6}$. Their values are~\cite{Yan:1992gz,Jiang:2014ena,Jiang:2015xqa}
\begin{eqnarray}
&&g_1=0.98,~~g_2=-\sqrt{\frac{3}{8}}g_1=-0.60,~~ g_3=\frac{\sqrt{3}}{2}g_1=0.85,\nonumber \\
&&g_4=-\sqrt{3}g_2=1.04,~~ g_5=-\frac{3}{2}g_1=-1.47,~~ g_6=0,
\end{eqnarray}
where $g_{2,4}$ are calculated through the strong decay widths of the charmed baryons and others are obtained through the quark model. In Table~\ref{63reults}, one obtains 
\begin{eqnarray}
\Gamma(\Xi_c^{'0} \rightarrow \Xi^0_c\gamma)= 0.46~\text{keV},\nonumber \\
 \Gamma(\Xi_c^{*0} \rightarrow \Xi^0_c \gamma)=1.84 ~\text{keV}.
\end{eqnarray}
The above results are independent of the inputs from the  lattice QCD simulations. For the neutral decay channel $ \Xi_c^{*0} \rightarrow \Xi^0_c \gamma$, the E2 transition decay width is only $1.6$ eV. The E2 transition is very strongly suppressed compared with the M1 transition.

\subsection{The radiative decay width from the spin-$3\over 2$ sextet to the spin-$1\over 2$ sextet }
In the heavy quark limit, the average mass differences are 
\begin{eqnarray}
\delta_1=\delta_2=127~\text{MeV}, ~~~\delta_3=0~\text{MeV}.
\end{eqnarray}
The mass difference between the antitriplet and sextet does not vanish in the heavy quark symmetry limit. This will impact the convergence of the numerical results~\cite{Wang:2018gpl}. Thus, we do not consider the contributions of the intermediate antitriplet states in the loops in the numerical analysis. 
Since $M_{-}=\delta_3$ vanishes in the heavy quark limit, the $G_2$ does not contribute to the $G_{M1}$. The $G_{E2}$ vanishes according to Eq.~(\ref{tr13}).  Then the $m_{1}$ and $\tilde m_{1}$ do not appear in the analytical expressions. The LECs are reduced to $\kappa_{1}$, $\kappa_{3}$, $\kappa_{4}$, $\kappa_{7}$ and $\kappa_8$.

 In Refs.~\cite{Wang:2018gpl,Meng:2018gan}, we decomposed the magnetic moments of the heavy baryons into the contributions of the light and heavy quarks.~We selected the average value $\mu_c=0.21\mu_N$ from the lattice QCD simulation as the magnetic moment of the charm quark. The heavy quark contributions to the magnetic moments of the antitriplet, the spin-$1\over 2$ and spin-$3\over 2$ sextets are $0.21\mu_N$, $-0.07\mu_N$ and $0.21\mu_N$, respectively. For the transition $B^\mu_{6^*}\rightarrow B_6 \gamma$, we use the $G^c_{M1}(\Omega^*_c\rightarrow \Omega_c\gamma)=-0.15$ in Ref.~\cite{Aubert:2006je} as the contribution of the charm quark. Then, we extract the contribution in the light quark sector and fit them order by order up to NNLO. The numerical results for the magnetic dipole form factors, the decay widths and the (transition) magnetic moments are listed in Table~\ref{GM1}. The chiral expansion works well. The chiral corrections at NLO and NNLO to the (transition) magnetic moments cancel with each other in most channels. This helps to guarantee that the total results are mainly from the leading order. 

\begin{table}
\caption{The magnetic dipole form factor, transition magnetic moment and the decay width for the radiative transition from the spin-$3\over 2$ to the spin-$1\over 2$ sextet. The second to the forth columns represent contributions from the light quarks order by oder.  The ``Light'' and ``Heavy'' represent the contributions from the light and heavy quarks, respectively. The sum of them are the total $G_{M1}$ from factor. ``$...$'' denotes that there is no corresponding data in the lattice QCD simulations.}\label{GM1}
\begin{tabular}{l|cccccc|c|c|c}
\toprule[1pt] \toprule[1pt] 
%\multicolumn{9}{|c|}{The $G_{M1}$ for spin-$\frac{3}{2}$ sextet to spin-$\frac{1}{2}$
%sextet } \tabularnewline
%\hline 
\multirow{2}{*}{Channel} & \multicolumn{7}{c|}{$G_{M1}$} & \multirow{2}{*}{$\mu_{6^*\rightarrow 6}$($\mu_N$)} & \multirow{2}{*}{$\Gamma (\text{keV})$}\tabularnewline
\cline{2-8} 

 & $\mathcal{O}(p)$ & $\mathcal{O}(p^{2})$ & $\mathcal{O}(p^{3})$ & Light  & Heavy & Total & lattice QCD \cite{Bahtiyar:2015sga} & \tabularnewline
\midrule[1pt] 
$\Sigma_{c}^{*++}\rightarrow\Sigma_{c}^{++}\gamma$ & 4.36 & -1.69 & 0.90 & 3.57 & $-0.15$ & $3.43$ & $...$ & $1.07$ & $1.20$\tabularnewline
 $\Sigma_{c}^{*+}\rightarrow\Sigma_{c}^{+}\gamma$ & 1.09 & -0.60 & 0.27 & 0.76 & $-0.15$ & 0.61 & $...$ & $0.19$ & $0.04$\tabularnewline
 $\Sigma_{c}^{*0}\rightarrow\Sigma_{c}^{0}\gamma$ & -2.18 & 0.49 & -0.37 & -2.06 & $-0.15$ & -2.20 & $...$ & $-0.69$ & $0.49$ \tabularnewline
 $\Xi_{c}^{*+}\rightarrow\Xi_{c}^{+'}\gamma$ & 1.15 & -0.26 & 0.04 & 0.92 & $-0.15$ & 0.77 & $...$ & $0.23$ & $0.07$ \tabularnewline  
$\Xi_{c}^{*0}\rightarrow\Xi_{c}^{0'}\gamma$ & -2.29 & 0.89 & -0.45 & -1.85 & $-0.15$ & -2.00 & $...$ & $-0.59$ & $0.42$\tabularnewline
 $\Omega_{c}^{*0}\rightarrow\Omega_{c}^{0}\gamma$ & -2.39 & 1.31 & -0.48 & -1.56 & $-0.15$ & $-1.71$ & $-0.816 $ & $-0.49$ & $0.32$\tabularnewline
\bottomrule[1pt] \bottomrule[1pt] 
\end{tabular}
\end{table}

\begin{table}
 \caption{The values of the LECs.}\label{lecs}
\begin{tabular}{cc|cc|cc|cc}
\toprule[1pt] \toprule[1pt] 
LECs & Value & LECs & Value & LECs & Value & LECs & Value\tabularnewline
\midrule[1pt] 
$\kappa_{1}$ & $-1.08$ & $f_{2}$ & $-0.48$ & $f_{4}$ & $-6.60$ & & \tabularnewline

$\kappa_{2}$ & $-0.83$ & $d_{5}$ & $2.87$ & $d_{8}$ & $2.15$ & $f_{3}$ & $-2.48$\tabularnewline
$\kappa_{3}$ & $0$ & $d_{6}$ & $0$ & $d_{9}$ & $0$ & $\tilde{f}_{3}$ & $0$\tabularnewline
$\kappa_{4}$ & $1.66$ & $a_{5}$ & $-1.91$ & $d_{4}$ & $1.10$ & $d_{7}$ & $1.66$\tabularnewline
$\kappa_{7}$ & $0.09$ & $l_{2}$ & $0.42$ & $s_{3}$ & $-0.24$ & $s_{8}$ & $0.36$\tabularnewline
$\kappa_{8}$ & $0$ & $l_{1}$ & $0$ & $s_{2}$ & $0$ & $s_{7}$ & $0$\tabularnewline
\bottomrule[1pt] \bottomrule[1pt] 
\end{tabular}
\end{table}

\section{The results for the bottom baryons}\label{sec6}
In this section, we extend the calculations to the singly bottom baryons.
The charge matrices of the bottom quark and bottom baryons are 
\begin{eqnarray}
\tilde Q_b=\text{diag}(-\frac{1}{3},-\frac{1}{3},-\frac{1}{3}), ~~~Q_{B}=\text{diag}(\frac{1}{2},-\frac{1}{2},-\frac{1}{2}). 
\end{eqnarray}

The (transition) magnetic moments and the radiative decay amplitudes of the singly heavy baryons can be divided as 
 \begin{eqnarray}
\mu=\mu^q+\mu^Q,~~~\mathcal M=\mathcal M^q+\mathcal M^Q,
 \end{eqnarray}
 where the superscripts ``$q$'' and ``$Q$'' denote the contributions from the light and heavy quarks, respectively. The Lagrangians and the LECs of the light quark sector are the same for the bottom and charmed baryons. For the heavy quark sector, one obtains the Lagrangians for the bottom baryons by replacing the $\tilde Q_c$ with $\tilde Q_b$ in the $\text{Tr}( f^+_{\mu\nu})$. 
 
In the heavy quark limit, the mass differences for the bottom baryon states are
 \begin{eqnarray}
 \delta_1=\delta_2= 157.39~ \text{MeV},~~~ \delta_3=0~ \text{MeV}.
 \end{eqnarray}

 For $B_{6}/B^{\mu}_{6^*}\rightarrow B_{\bar 3} \gamma$, one obtains 
 \begin{eqnarray}
 \Gamma(\Xi_{b}^{'-}\rightarrow\text{\ensuremath{\Xi}}_{b}^{-}\gamma)=1.0~\text{keV},~~\Gamma(\Xi_{b}^{*-}\rightarrow\text{\ensuremath{\Xi}}_{b}^{-}\gamma)=1.4~\text{keV}.
 \end{eqnarray}
The $\Gamma(\Xi_{b}^{*-}\rightarrow\text{\ensuremath{\Xi}}_{b}^{-}\gamma)$ is also mainly from the M1 transition, and the E2 decay width is only $0.20$ eV.

 \begin{table}
 \caption{The transition magnetic moment and the decay width for the bottom sextet to the antitriplet. $\mu$ is in the unit of the nuclear magneton.}\label{63breults}
 \begin{tabular}{l|ccc|c}
\toprule[1pt] \toprule[1pt] 
\multirow{2}{*}{ Channel} & \multicolumn{3}{c|}{$\mu$} & \multirow{2}{*}{$\Gamma$(keV)}\tabularnewline
\cline{2-4} 
& $\mathcal{O}(p)$ & \multicolumn{1}{c}{$\mathcal{O}(p^{2})$} & Total & \tabularnewline
 \midrule[1pt]
  
$\Sigma_{b}^{0}\rightarrow\Lambda_{b}^{0}\gamma$ & -2.70 & 1.33 & -1.37 & 108.0  \tabularnewline
 
$\Xi_{b}^{'0}\rightarrow\text{\ensuremath{\Xi}}_{b}^{0}\gamma$ & -2.70 & 1.95 & -0.75 & 13.0 \tabularnewline
 $\Xi_{b}^{'-}\rightarrow\text{\ensuremath{\Xi}}_{b}^{-}\gamma$ & 0 & 0.21 & 0.21 & 1.0 \tabularnewline
 \midrule[1pt]
$\Sigma_{b}^{*0}\rightarrow\Lambda_{b}^{0}\gamma$ & 3.85 & -1.89 & $1.96$ & 142.1\tabularnewline
  
$\Xi_{b}^{*0}\rightarrow\text{\ensuremath{\Xi}}_{b}^{0}\gamma$ & 3.84 & -2.78 & $1.06$ & 17.2\tabularnewline
  
$\Xi_{b}^{*-}\rightarrow\text{\ensuremath{\Xi}}_{b}^{'-}\gamma$ & $0$ & $-0.30$ & $-0.30$ & $1.4$ \tabularnewline

\bottomrule[1pt] \bottomrule[1pt] 
\end{tabular}
\end{table}

%the mass difference of the initial and final states is $0$ in the heavy quark limit. 
For the radiative decays $B^\mu_{6^*}\rightarrow B_6\gamma$, we use the predictions from the quark model to estimate the contributions from the bottom quarks~\cite{Meng:2018gan}. The transition magnetic moments and the radiative decay widths are listed in Table~\ref{mub66}.

\begin{table}
 \caption{The transition magnetic moment and the decay width for the transition $B^\mu_{6^*}\rightarrow B_6\gamma$. }\label{mub66}
\begin{tabular}{lccc|c}
\toprule[1pt] \toprule[1pt] 
$\mu_{6^{*}\rightarrow6}$ & Light & Heavy & Total & $\Gamma$(eV) \tabularnewline
\midrule[1pt] 
$\Sigma_{b}^{*+}\rightarrow\Sigma_{b}^{+}\gamma$ & $1.11$ & $0.06$ & $1.17$ & $50$\tabularnewline
  
$\Sigma_{b}^{*0}\rightarrow\Sigma_{b}^{0}\gamma$ & $0.24$ & $0.06$ & $0.30$ & $3.0$\tabularnewline
  
$\Sigma_{b}^{*-}\rightarrow\Sigma_{b}^{-}\gamma$ & $-0.63$ & $0.06$ & $-0.58$ & $10.3$\tabularnewline
  
$\Xi_{b}^{*0}\rightarrow\Xi_{b}^{'0}\gamma$ & $0.27$ & $0.06$ & $0.33$ & $1.5$\tabularnewline
  
$\Xi_{b}^{*-}\rightarrow\Xi_{b}^{'-}\gamma$ & $-0.54$ & $0.06$ & $-0.49$ & $8.2$\tabularnewline
 $\Omega_{b}^{*-}\rightarrow\Omega_{b}^{-}\gamma$ & $-0.44$ & $0.06$ & $-0.38$ & $30.6$\tabularnewline
\bottomrule[1pt] \bottomrule[1pt] 
\end{tabular}
\end{table}

\section{Summary}\label{sec7}
In this work, we calculate the radiative decay amplitudes and the transition magnetic moments for the singly heavy baryons. We derive their analytical expressions up to the next-to-next-to-leading order in the framework of the HBChPT. The expressions contain many LECs. Most of them also contributed to the magnetic moments. Thus, we perform the numerical analysis for the magnetic moments and the decay amplitudes of the singly heavy baryons simultaneously with a set of unified LECs. The heavy baryons have the heavy quark symmetry in the heavy quark limit. This helps to reduce the number of the independent LECs.

For the decays $B_6\rightarrow B_{\bar 3}\gamma$ and  $B^\mu_{6^*}\rightarrow B_{\bar 3}\gamma$, we calculate the numerical results up to the next-to-leading order. Due to the U-spin symmetry, the tree diagrams do not contribute to the transitions $\Xi_c^{'0} \rightarrow \Xi^0_c\gamma$ and $\Xi_c^{*0} \rightarrow \Xi^0_c \gamma$. Their decay widths totally arise from the chiral corrections, which does not involve unknown LECs up to NLO. For $\Xi_c^{*0} \rightarrow \Xi^0_c \gamma$, the E2 transition is suppressed. The above conclusions also hold for the radiative decays $\Xi_b^{'-} \rightarrow \Xi^-_b\gamma$ and $\Xi_b^{*-} \rightarrow \Xi^-_b \gamma$. 

For the radiative decays $B^\mu_{6^*}\rightarrow B_6\gamma$, we calculate numerical results of the decay widths up to the next-to-next-to-leading order. In the process, we do not include the antitriplet states as the intermediate states in the loops. We use the magnetic moments  of the charmed baryons from the lattice QCD simulations are treated as input and predict the transition magnetic moments and the decay widths. 

We extend the calculations to the bottom baryons. The light quark contributions are the same as those in the charmed baryon sector. The heavy quark contributions are estimated using the quark model.

In Tables~\ref{dw} and~\ref{dwb}, we list our numerical results for the radiative decay widths in the charmed and bottom baryon sectors, respectively. We compare them with the results calculated using the 
lattice QCD~\cite{Bahtiyar:2015sga,Bahtiyar:2016dom}, the extent bag model~\cite{Simonis:2018rld}, the light cone QCD sum rule~\cite{Aliev:2009jt,Aliev:2014bma,Aliev:2016xvq}, the heavy hadron chiral perturbation theory (HHChPT)~\cite{Cheng:1993kp,Banuls:1999br}, the HBChPT~\cite{Jiang:2015xqa} and the quark model~\cite{Ivanov:1999bk}. For the radiative decays $B_6\rightarrow B_{\bar 3}\gamma$ and $B^{\mu}_{6^*}\rightarrow B_{\bar 3}\gamma$, our numerical results are consistent with those from other frameworks. For the radiative decay $B^{\mu}_{6^*}\rightarrow B_{\bar 6}\gamma$, we have estimated the LECs by adopting four magnetic moments  from the lattice QCD simulations as input, which are smaller than those of other models~\cite{Wang:2018gpl,Meng:2018gan}. Since the decay width is proportional to the square of the multipole form factor, the inputs from the lattice QCD may lead to smaller decay widths. 
%For instance, the $\Gamma(\Xi_{c}^{'0}\rightarrow\text{\ensuremath{\Xi}}_{c}^{0}\gamma)$ in our calculation, which is independent of the inputs, is much larger than that from the lattice QCD simulations.

 In the future, with more data from the experiment and the lattice QCD, we can update our numerical results using the analytical expressions.~We expect the analytical expressions may be helpful for the extroplation of the lattice QCD simulation. Hopefully, our numerical results will be helpful to the experimental search of the radiative decays of the heavy baryons at LHCb, Belle II and BESIII.
\begin{table}
 \caption{The decay widths of the charmed baryon transitions from different frameworks, the 
lattice QCD~\cite{Bahtiyar:2015sga,Bahtiyar:2016dom}, the extent bag model~\cite{Simonis:2018rld}, the light cone QCD sum rule~\cite{Aliev:2009jt,Aliev:2014bma,Aliev:2016xvq}, the heavy hadron chiral perturbation theory (HHChPT)~\cite{Cheng:1993kp,Banuls:1999br}, the HBChPT~\cite{Jiang:2015xqa} and the quark model~\cite{Ivanov:1999bk}. }\label{dw}
\begin{tabular}{l|cccccccc}
\toprule[1pt] \toprule[1pt] 
$\Gamma$ (keV) & This work &~\cite{Bahtiyar:2015sga,Bahtiyar:2016dom} &~\cite{Simonis:2018rld} &~\cite{Aliev:2009jt,Aliev:2014bma,Aliev:2016xvq} &~\cite{Cheng:1993kp} & ~\cite{Banuls:1999br} &~\cite{Jiang:2015xqa} &~\cite{Ivanov:1999bk}\tabularnewline
\midrule[1pt] 
$\Sigma_{c}^{+}\rightarrow\Lambda_{c}^{+}\gamma$ & $65.6 $ & $...$ & $74.1$ & $50(17)$ & $46$ & ... & $164$ & $60.7\pm1.5$\tabularnewline
 
$\Xi_{c}^{'+}\rightarrow\text{\ensuremath{\Xi}}_{c}^{+}\gamma$ & $5.43$ & $5.468(1.500)$ & $17.3$ & $8.5(2.5)$ & $1.3$ & ... & $54.3$ & $12.7\pm1.5$\tabularnewline

$\Xi_{c}^{'0}\rightarrow\text{\ensuremath{\Xi}}_{c}^{0}\gamma$ & $0.46 $ & $0.002(4)$ & $0.185$ & $0.27(6)$ & $0.04$ & $1.2\pm0.7$ & $0.02$ & $0.17\pm0.02$\tabularnewline
\midrule[1pt]
{  $ {\Sigma_{c}^{*+}\rightarrow\Lambda_{c}^{+}\gamma}$} & $161.8$ &$...$  & 190 & $130(45)$ &$...$ &$...$ & 893  &$151\pm 4$ \tabularnewline
 {  $ {\Xi_{c}^{*+}\rightarrow\text{\ensuremath{\Xi}}_{c}^{+}\gamma}$} & $21.6$ &$...$  &72.7 & $52(25)$ &$...$ &$...$ & 502& $54 \pm 3$ \tabularnewline
$\Xi_{c}^{*0}\rightarrow\text{\ensuremath{\Xi}}_{c}^{0}\gamma$ & $1.84$ & $...$ & $0.745$ & $0.66(32)$ & ... & $5.1\pm2.7$ & $0.36$ & $0.68\pm0.04$\tabularnewline
\midrule[1pt]
$\Sigma_{c}^{*++}\rightarrow\Sigma_{c}^{++}\gamma$ & $1.20$ & $...$ & $1.96$ & $2.65(1.20)$ &... & ... & $11.6$ & $...$\tabularnewline

$\Sigma_{c}^{*+}\rightarrow\Sigma_{c}^{+}\gamma$ & $0.04$ & $...$ & $0.011$ & $0.46(16)$ & ...& ... & $0.85$ & $0.14\pm0.004$\tabularnewline

$\Sigma_{c}^{*0}\rightarrow\Sigma_{c}^{0}\gamma$ & $0.49$ & $...$ & $1.41$ & $0.08(3)$ & ...& ... & $2.92$ & ...\tabularnewline

$\Xi_{c}^{*+}\rightarrow\Xi_{c}^{'+}\gamma$ & $0.07$ & $...$ & $0.063$ & $0.274$ & ...& ... & $1.10$ & $...$\tabularnewline

$\Xi_{c}^{*0}\rightarrow\Xi_{c}^{'0}\gamma$ & $0.42$ & $...$ & $1.33$ & $2.14$ & ... & ... & $3.83$ & $...$\tabularnewline

$\Omega_{c}^{*0}\rightarrow\Omega_{c}^{0}\gamma$ & $0.32$ & $0.074(8)$ & $1.13$ & $0.932$ &... & ... & $4.82$ & ...\tabularnewline
\bottomrule[1pt] \bottomrule[1pt] 
\end{tabular}
 \end{table}
 
 \begin{table}
 \caption{The decay widths of the bottom baryon transitions from different frameworks, the extent bag model~\cite{Simonis:2018rld}, the light cone QCD sum rule~\cite{Aliev:2009jt,Aliev:2014bma,Aliev:2016xvq}, the HHChPT~\cite{Banuls:1999br} and the HBChPT~\cite{Jiang:2015xqa} . }\label{dwb}
 \begin{tabular}{l|ccccc}
\toprule[1pt] \toprule[1pt] 
$\Gamma$ (keV) & This work& ~\cite{Simonis:2018rld} & ~\cite{Aliev:2009jt,Aliev:2014bma,Aliev:2016xvq} &~\cite{Banuls:1999br} & \cite{Jiang:2015xqa}\tabularnewline
 \midrule[1pt] 
$\Sigma_{b}^{0}\rightarrow\Lambda_{b}^{0}\gamma$ & $108.0 $ & $116$ & $152(60)$ & ... &288\tabularnewline

$\Xi_{b}^{'0}\rightarrow\text{\ensuremath{\Xi}}_{b}^{0}\gamma$ & $13.0$ & $36.4$ & $47(21)$ & ... & ...\tabularnewline

$\Xi_{b}^{'-}\rightarrow\text{\ensuremath{\Xi}}_{b}^{-}\gamma$ & $1.0$ & $0.357$ & $3.3(1.3)$ & $3.1\pm1.8$ & ...\tabularnewline
\midrule[1pt]

{ $\Sigma_{b}^{0}\rightarrow\Lambda_{b}^{*0}\gamma$} & 142.1 & 158  & 114 (45) &$...$ & 435\tabularnewline
  
{$\Xi_{b}^{*0}\rightarrow\text{\ensuremath{\Xi}}_{b}^{0}\gamma$} & 17.2 &55.3  &135(65) &$...$ &136\tabularnewline

$\Xi_{b}^{*-}\rightarrow\text{\ensuremath{\Xi}}_{b}^{-}\gamma$ & $1.4 $ & $0.536$ & $1.50(75)$ & $4.2\pm2.4$ &1.87\tabularnewline
\midrule[1pt]

$\Sigma_{b}^{*+}\rightarrow\Sigma_{b}^{+}\gamma$ & $0.05$ & $0.11$ & $0.46(22)$ & ...& 0.6\tabularnewline

$\Sigma_{b}^{*0}\rightarrow\Sigma_{b}^{0}\gamma$ & $3.0\times10^{-3}$ & $8.3\times10^{-3}$ & $0.028(16)$ & ... &0.05\tabularnewline

$\Sigma_{b}^{*-}\rightarrow\Sigma_{b}^{-}\gamma$ & $0.013$ & $0.0192$ & $0.11(6)$ & ... &0.08\tabularnewline

$\Xi_{b}^{*0}\rightarrow\Xi_{b}^{'0}\gamma$ & $1.5\times 10^{-3}$ & $0.0105$ & $0.131$ & ... & ...\tabularnewline

$\Xi_{b}^{*-}\rightarrow\Xi_{b}^{'-}\gamma$ & $8.2 \times10^{-3}$ & $0.0136$ & $0.303$ & ...& ...\tabularnewline

$\Omega_{b}^{*-}\rightarrow\Omega_{b}^{-}\gamma$ & $0.031$ & $9.1\times10^{-3}$ & $0.092$ & ...& ... \tabularnewline

\bottomrule[1pt] \bottomrule[1pt] 
\end{tabular}
 \end{table}

\section*{Acknowledgements}

 This project is supported by the National
Natural Science Foundation of China under Grants 11575008,
11621131001 and National Key Basic Research Program of
China(2015CB856700).

\begin{appendix}

\section{Magnetic moments of spin-$\frac{1}{2}$ and spin-$\frac{3}{2}$ sextets}\label{app:mm}
In this section, we give the magnetic moments of spin-$\frac{1}{2}$ and spin-$\frac{3}{2}$ sextets in Tables~\ref{muu:charm} and~\ref{mm:bottom}.
\begin{table}
\caption{The magnetic moments of the spin-$\frac{1}{2}$ and spin-$\frac{3}{2}$ singly charmed sextet.}\label{muu:charm}
\begin{tabular}{c|cccccc|c}
\toprule[1pt]\toprule[1pt]
 & $\mathcal{O}(p)$ & $\mathcal{O}(p^{2})$ & $\mathcal{O}(p^{3})$ & Light & Heavy & Total & lattice QCD\tabularnewline
 \midrule[1pt] 

$\mu_{\Sigma_{c}^{++}}^{\ddagger}$ & 1.91 & -0.74 & 0.39 & 1.57 & -0.07 & 1.50 & $1.499(202)$\tabularnewline
 $\mu_{\Sigma_{c}^{+}}$ & 0.48 & -0.26 & 0.12 & 0.33 & -0.07 & 0.26 & $...$\tabularnewline
 $\mu_{\Sigma_{c}^{0}}$ & -0.96 & 0.22 & -0.16 & -0.90 & -0.07 & -0.97 & $-0.875(103)$\tabularnewline
 $\mu_{\Xi_{c}^{+'}}^{\ddagger}$ & 0.48 & -0.11 & 0.01 & 0.39 & -0.07 & 0.32 & $0.315(141)$ \tabularnewline
  
$\mu_{\Xi_{c}^{0}}$ & -0.96 & 0.37 & -0.19 & -0.77 & -0.07 & -0.84 & $-0.599(71)$ \tabularnewline
 $\mu_{\Omega_{c}^{0}}^{\ddagger}$ & -0.96 & 0.52 & -0.19 & -0.62 & -0.07 & -0.69 & $-0.688(31)$ \tabularnewline
\midrule[1pt]
$\mu_{\Sigma_{c}^{*++}}$ & $2.87$ & $-1.11$ & $0.59$ & $2.35$ & $0.21$ & $2.56$ & $...$\tabularnewline
 
$\mu_{\Sigma_{c}^{*+}}$ & $0.72$ & $-0.39$ & $0.17$ & $0.50$ & $0.21$ & $0.71$ & $...$\tabularnewline
 $\mu_{\Sigma_{c}^{*0}}$ & $-1.43$ & $0.32$ & $-0.24$ & $-1.35$ & $0.21$ & $-1.14$ & $...$\tabularnewline
 $\mu_{\Xi_{c}^{*+}}$ & $0.72$ & $-0.16$ & $0.02$ & $0.58$ & $0.21$ & $0.79$ & $...$\tabularnewline
  
$\mu_{\Xi_{c}^{*0}}$ & $-1.43$ & $0.55$ & $-0.28$ & $-1.16$ & $0.21$ & $-0.95$ & $...$\tabularnewline
  
$\mu_{\Omega_{c}^{*0}}^{\ddagger}$ & $-1.43$ & $0.78$ & $-0.29$ & $-0.94$ & $0.21$ & $-0.73$ & $-0.730(23)$\tabularnewline
\bottomrule[1pt] \bottomrule[1pt] 
\end{tabular}
\end{table}

\begin{table}
\caption{The magnetic moments of the spin-$\frac{1}{2}$ and spin-$\frac{3}{2}$ singly bottom sextet.}\label{mm:bottom}
\begin{tabular}{lccc|cccc}
\toprule[1pt] \toprule[1pt] 
 & Light & Heavy & Total & & Light & Heavy & Total\tabularnewline
\midrule[1pt] 

$\mu_{\Sigma_{b}^{+}}$ & $1.57$ & $-0.02$ & $1.55$ & $\mu_{\Sigma_{b}^{*+}}$ & $2.35$ & $-0.06$ & $2.29$\tabularnewline
 $\mu_{\Sigma_{b}^{0}}$ & $0.33$ & $-0.02$ & $0.31$ & $\mu_{\Sigma_{b}^{*0}}$ & $0.50$ & $-0.06$ & $0.44$\tabularnewline
 $\mu_{\Sigma_{b}^{-}}$ & $-0.90$ & $-0.02$ & $-0.92$ & $\mu_{\Sigma_{b}^{*-}}$ & $-1.35$ & $-0.06$ & $-1.41$\tabularnewline
 $\mu_{\Xi_{b}^{'0}}$ & $0.39$ & $-0.02$ & $0.37$ & $\mu_{\Xi_{b}^{*0}}$ & $0.58$ & $-0.06$ & $0.51$\tabularnewline
 $\mu_{\Xi_{b}^{'-}}$ & $-0.77$ & $-0.02$ & $-0.79$ & $\mu_{\Xi_{b}^{*-}}$ & $-1.16$ & $-0.06$ & $-1.22$\tabularnewline
 $\mu_{\Omega_{b}^{-}}$ & $-0.62$ & $-0.02$ & $-0.64$ & $\mu_{\Omega_{b}^{*-}}$ & $-0.94$ & $-0.06$ & $-1.00$\tabularnewline
\bottomrule[1pt]\bottomrule[1pt]
\end{tabular}
\end{table}

\section{Quark model results}\label{QM}
We calculate the transition magnetic moments of the charmed baryons in the quark model. For the radiative decays $B_6\rightarrow B_{\bar 3}\gamma$ and $B^\mu_{6^*}\rightarrow B_{\bar 3}\gamma$, the results are 
\begin{eqnarray}
&&\mu({\Sigma_{c}^{+}\rightarrow\Lambda_{c}^{+}\gamma})=-\frac{1}{\sqrt{3}}(\mu_{u}-\mu_{d}),\nonumber\\
&&{\mu_({\Xi_{c}^{'+}\rightarrow\Xi_{c}^{+}\gamma}})=-\frac{1}{\sqrt{3}}(\mu_{u}-\mu_{s}),\nonumber\\
&&{\mu_({\Xi_{c}^{'0}\rightarrow\Xi_{c}^{0}\gamma}})=-\frac{1}{\sqrt{3}}(\mu_{d}-\mu_{s}),\nonumber\\
&&\mu({\Sigma_{c}^{*+}\rightarrow\Lambda_{c}^{+}\gamma})= \ensuremath{\frac{2}{\sqrt{6}}(\mu_{u}-\mu_{d})},\nonumber\\
&&{\mu_({\Xi_{c}^{*+}\rightarrow\Xi_{c}^{+}\gamma}})=\frac{2}{\sqrt{6}}(\mu_{u}-\mu_{s}),\nonumber\\
&&{\mu_({\Xi_{c}^{*0}\rightarrow\Xi_{c}^{0}\gamma}})=\frac{2}{\sqrt{6}}(\mu_{d}-\mu_{s}),
\end{eqnarray}
where we use $\mu_{u,d,s}$ and $\mu_{c}$ to denote the magnetic moments of the light and heavy quarks, respectively. We find that the heavy quarks do not contribute to the radiative decays from the sextet to the antitriplet. The contributions of two light quarks are opposite to each other.

For the decay $B^\mu_{6^*}\rightarrow B_6\gamma$, one obtains,
\begin{eqnarray}
&&{\mu({\Sigma_{c}^{*++}\rightarrow\Sigma_{c}^{++}}}\gamma)=\frac{\sqrt{2}}{3}(2\mu_{\mu}-2\mu_{c}),\nonumber\\
&&{\mu({\Sigma_{c}^{*+}\rightarrow\Sigma_{c}^{+}}}\gamma)=\ensuremath{\frac{\sqrt{2}}{3}(\mu_{u}+\mu_{d}-2\mu_{c})},\nonumber\\
&&\mu({\Sigma_{c}^{*0}\rightarrow\Sigma_{c}^{0}}\gamma)=\ensuremath{\frac{\sqrt{2}}{3}(2\mu_{d}-2\mu_{c})},\nonumber\\
&&{\mu({\Xi_{c}^{*'+}\rightarrow\Xi_{c}^{'+}}}\gamma)=\ensuremath{\frac{\sqrt{2}}{3}(\mu_{u}+\mu_{s}-2\mu_{c})},\nonumber\\
&&{\mu({\Xi_{c}^{*0}\rightarrow\Xi_{c}^{'0}}}\gamma)=\frac{\sqrt{2}}{3}(\mu_{s}+\mu_{d}-2\mu_{c}),\nonumber\\
&&{\mu({\Omega_{c}^{*0}\rightarrow\Omega_{c}^{0}}}\gamma)=\frac{\sqrt{2}}{3}(2\mu_{s}-2\mu_{c}).
\end{eqnarray}
Both the light and heavy quarks contribute to the transition magnetic moments. 

\section{The loop integrals}\label{loop}
In this section, we list the loop integrals involved in this work. 
\begin{eqnarray}
\Delta=i\int\frac{d^{d}\lambda^{4-d}}{(2\pi)^{d}}\frac{1}{l^{2}-m^{2}+i\epsilon}=2m^{2}\left(L(\lambda)+\frac{1}{32\pi^{2}}\text{ln}\frac{m^{2}}{\lambda^{2}}\right),
\end{eqnarray}
where 
\begin{eqnarray}
L(\lambda)=\frac{\lambda^{d-4}}{16\pi^{2}}\left[\frac{1}{d-4}-\frac{1}{2}\left(\text{ln}(4\pi)+1+\Gamma'(1)\right)\right].
\end{eqnarray}

\begin{eqnarray}
I_{0}(q^{2}) & =&i\int\frac{d^{d}\lambda^{4-d}}{(2\pi)^{d}}\frac{1}{\left(l^{2}-m^{2}+i\epsilon\right)\left((l+q)^{2}-m^{2}+i\epsilon\right)},\\
I_{0}(q^{2}) & =&\begin{cases}
-\frac{1}{16\pi^{2}}\left(1-\text{ln}\frac{m^{2}}{\lambda^{2}}-r\text{ln}|\frac{1+r}{1-r}|\right)+2L(\lambda) & (q^{2}<0),\\
-\frac{1}{16\pi^{2}}\left(1-\text{ln}\frac{m^{2}}{\lambda^{2}}-2r\text{arctan}\frac{1}{r}\right)+2L(\lambda) & (0<q^{2}<4m^{2}),\\
-\frac{1}{16\pi^{2}}\left(1-\text{ln}\frac{m^{2}}{\lambda^{2}}-r\text{ln}|\frac{1+r}{1-r}|+i\pi r\right)+2L(\lambda) & (q^{2}>4m^{2}).
\end{cases}
\end{eqnarray}

where $r=\sqrt{|1-\frac{4m^{2}}{q^{2}}|}$.

\begin{eqnarray}
i\int\frac{d^{d}l\,\lambda^{4-d}}{(2\pi)^{d}}\frac{[1,l_{\alpha},l_{\alpha}l_{\beta}]}{(l^{2}-m^{2}+i\epsilon)(\omega+v\cdot l+i\epsilon)}=[J_{0}(\omega),v_{\alpha}J_{1}(\omega),g_{\alpha\beta}J_{2}(\omega)+v_{\alpha}v_{\beta}J_{3}(\omega)].
\end{eqnarray}

\begin{eqnarray}
J_{0}(\omega)=\begin{cases}
{\displaystyle \frac{-\omega}{8\pi^{2}}(1-\ln\frac{m^{2}}{\lambda^{2}})+\frac{\sqrt{\omega^{2}-m^{2}}}{4\pi^{2}}({\rm \text{arccosh}}\frac{\omega}{m}-i\pi)+4\omega L(\lambda)} & (\omega>m),\\
{\displaystyle \frac{-\omega}{8\pi^{2}}(1-\ln\frac{m^{2}}{\lambda^{2}})+\frac{\sqrt{m^{2}-\omega^{2}}}{4\pi^{2}}\arccos\frac{-\omega}{m}+4\omega L(\lambda)} & (\omega^{2}<m^{2}),\\
{\displaystyle \frac{-\omega}{8\pi^{2}}(1-\ln\frac{m^{2}}{\lambda^{2}})-\frac{\sqrt{\omega^{2}-m^{2}}}{4\pi^{2}}{\rm \text{arccosh}}\frac{-\omega}{m}+4\omega L(\lambda)} & (\omega<-m).
\end{cases}
\end{eqnarray}

\begin{eqnarray}
&&J_{1}(\omega)= -\omega J_{0}(\omega)+\Delta.\\
&&J_{2}(\omega)= \frac{1}{d-1}[(m^{2}-\omega^{2})J_{0}(\omega)+\omega\Delta].\\
&&J_{3}(\omega)= -\omega J_{1}(\omega)-J_{2}(\omega).\\
&&\Lambda_2(\omega_1,\omega_2) \equiv \frac{J_2(\omega_1)-J_2(\omega_2)}{\omega_2-\omega_1}.
\end{eqnarray}

The loops that contain a heavy baryon and two meson propagators can be expressed as
\begin{eqnarray}
i\int\frac{d^{d}l\,\lambda^{4-d}}{(2\pi)^{d}}\frac{[1,l_{\alpha},l_{\alpha}l_{\beta},l_{\nu}l_{\alpha}l_{\beta}]}{(l^{2}-m^{2}+i\epsilon)((l+q)^{2}-m^{2}+i\epsilon)(\omega+v\cdot l+i\epsilon)}=[L_{0}(\omega),L_{\alpha},L_{\alpha\beta},L_{\nu\alpha\beta}],
\end{eqnarray}
with $\beta=\omega-v\cdot q$ and $v\cdot q>0$.
\begin{eqnarray}
L_{0}(\omega)=\begin{cases}
{\displaystyle \frac{1}{8\pi^{2}v\cdot q}\left\{ \frac{1}{2}\left[\left({\rm arccosh}\frac{\beta}{m}\right)^{2}-\left({\rm arccosh}\frac{\omega}{m}\right)^{2}\right]-i\pi\ln\frac{\sqrt{\beta^{2}-m^{2}}+\beta}{\sqrt{\omega^{2}-m^{2}}+\omega}\right\} } , (\beta>m)\\
{\displaystyle \frac{1}{16\pi^{2}v\cdot q}\left[\left({\rm arccos}\frac{-\omega}{m}\right)^{2}-\left({\rm arccos}\frac{-\beta}{m}\right)^{2}\right]},\,\, \, (\beta^{2}<m^{2})\\
{\displaystyle \frac{1}{16\pi^{2}v\cdot q}\left[\left({\rm arccosh}\frac{-\beta}{m}\right)^{2}-\left({\rm arccosh}\frac{-\omega}{m}\right)^{2}\right]}.\,\, \, (\beta<-m)
\end{cases}
\end{eqnarray}

\begin{eqnarray}
L_{\alpha\beta}&=&n_{1}^{\textrm{II}}g_{\alpha\beta}+n_{2}^{\textrm{II}}q_{\alpha}q_{\beta}+n_{3}^{\textrm{II}}v_{\alpha}v_{\beta}+n_{4}^{\textrm{II}}v_{\alpha}q_{\beta}+n_{5}^{\textrm{II}}q_{\alpha}v_{\beta},\\
L_{\nu\alpha\beta} & =&n_{1}^{\text{III}}q_{\nu}q_{\alpha}q_{\beta}+n_{2}^{\text{III}}q_{\nu}q_{\alpha}v_{\beta}+n_{3}^{\text{III}}q_{\nu}q_{\beta}v_{\alpha}+n_{4}^{\text{III}}q_{\alpha}q_{\beta}v_{\nu}\nonumber \\
 &~~&+n_{5}^{\text{III}}q_{\nu}g_{\alpha\beta}+n_{6}^{\text{III}}q_{\beta}g_{\nu\alpha}+n_{7}^{\text{III}}q_{\alpha}g_{\nu\beta}\nonumber \\ 
 &~~&+n_{8}^{\text{III}}q_{\nu}v_{\alpha}v_{\beta}+n_{9}^{\text{III}}q_{\alpha}v_{\nu}v_{\beta}+n_{10}^{\text{III}}q_{\beta}v_{\nu}v_{\alpha}\nonumber \\
 &~~&+n_{11}^{\text{III}}g_{\nu\beta}v_{\alpha}+n_{12}^{\text{III}}g_{\nu\alpha}v_{\beta}+n_{13}^{\text{III}}g_{\alpha\beta}v_{\nu}+n_{14}^{\text{III}}v_{\nu}v_{\alpha}v_{\beta}.
\end{eqnarray}
The explicit forms of the $n_{1}^{\text{II}}$, $n_{2}^{\text{III}}$ and so on are quite complex. We list their relations with some simple integrals. 
\begin{eqnarray}
n_{1}^{\text{II}}(\omega,\beta)&=&\frac{-\beta J_{0}\text{(\ensuremath{\beta})}+J_{0}(\omega)\omega+2L_{0}(\omega,\beta)m^{2}(\omega-\beta)}{2(d-2)(\omega-\beta)}.\\
n_{4}^{\text{II}}(\omega,\beta)&=&n_{5}^{\text{II}}(\omega,\beta)=\frac{J_{0}(\text{\ensuremath{\beta})}((d-2)\omega-\beta(d-3))-J_{0}(\omega)\omega+2L_{0}m^{2}(\beta-\omega)}{2(d-2)(\omega-\beta)^{2}}.\\
n_{2}^{\text{III}}(\omega,\beta)&=&-\frac{J_{0}(\text{\ensuremath{\beta}})\left[\beta^{2}\left(d^{2}-5d+6\right)+\left(d^{2}-3d+2\right)\omega^{2}-2\beta\left(d^{2}-4d+3\right)\omega+2(d-2)m^{2}\right]}{2(d-2)(d-1)(\omega-\beta)^{3}}\nonumber\\
&~~&-\frac{4m^{2}(\omega-\beta)\left[(d-2)I_{0}-(d-1)L_{0}(\omega,\beta)\omega\right]-2J_{0}(\omega)\left[(d-2)m^{2}+\omega^{2}\right]}{2(d-2)(d-1)(\omega-\beta)^{3}}.\\
n_{5}^{\text{III}}(\omega,\beta)&=&\frac{2m^{2}(\omega-\beta)\left[(d-2)I_{0}(q^{2})-(d-1)L_{0}(\omega,\beta)\omega\right]}{2(d-2)(d-1)(\omega-\beta)^{2}}\nonumber\\
&~~&+\frac{\text{\ensuremath{J_{0}(\beta)}}\left[(d-2)m^{2}+\beta^{2}-\beta(\beta+\omega)(d-1)\right]-J_{0}(\omega)\left[(d-2)m^{2}+\omega^{2}\right]}{2(d-2)(d-1)(\omega-\beta)^{2}}.
\end{eqnarray}
\end{appendix}

\end{document}